\documentclass[12pt, reqno]{amsart}
\usepackage{amsmath,amssymb,pdflscape,bm,xr}
\usepackage{amsfonts,natbib}
\usepackage{graphicx,epsfig,setspace}
\usepackage[margin=1in]{geometry}

\numberwithin{equation}{section}
\begin{document}

\newcommand{\cov}{\textnormal{Cov}}
\newcommand{\var}{\textnormal{Var}}
\newcommand{\diag}{\textnormal{diag}}
\newcommand{\plim}{\textnormal{plim}_n}
\newcommand{\dum}{1\hspace{-2.5pt}\textnormal{l}}
\newcommand{\ind}{\bot\hspace{-6pt}\bot}
\newcommand{\co}{\textnormal{co}}
\newcommand{\tr}{\textnormal{tr }}
\newcommand{\fsgn}{\textnormal{\footnotesize sgn}}
\newcommand{\sgn}{\textnormal{sgn}}
\newcommand{\fatb}{\mathbf{b}}
\newcommand{\fatp}{\mathbf{p}}
\newcommand{\trace}{\textnormal{trace}}

\newtheorem{dfn}{Definition}[section]
\newtheorem{rem}{Remark}[section]
\newtheorem{cor}{Corollary}[section]
\newtheorem{thm}{Theorem}[section]
\newtheorem{lem}{Lemma}[section]
\newtheorem{notn}{Notation}[section]
\newtheorem{con}{Condition}[section]
\newtheorem{prp}{Proposition}[section]
\newtheorem{pty}{Property}[section]
\newtheorem{ass}{Assumption}[section]
\newtheorem{ex}{Example}[section]
\newtheorem*{cst1}{Constraint S}
\newtheorem*{cst2}{Constraint U}
\newtheorem{qn}{Question}[section]


\onehalfspacing

\title[Multi-Way Clustering]{Bootstrap with Clustering in Two or More Dimensions}
\author[Konrad Menzel]{Konrad Menzel\\New York University}
\date{November 2016 - this version: December 2017. The author thanks Matias Cattaneo, Tim Christensen, Bryan Graham, and Valentin Verdier for useful comments and gratefully acknowledges support from the NSF (SES-1459686).}

\begin{abstract}
We propose a bootstrap procedure for data that may exhibit cluster dependence in two or more dimensions. We use insights from the theory of generalized U-statistics to analyze the large-sample properties of statistics that are sample averages from the observations pooled across clusters. The asymptotic distribution of these statistics may be non-standard if observations are dependent but uncorrelated within clusters. We show that there exists no procedure for estimating the limiting distribution of the sample mean under two-way clustering that achieves uniform consistency. However, we  propose (a) one bootstrap procedure that is adaptive and point-wise consistent for any fixed data-generating process (DGP), (b) an alternative procedure that is uniformly consistent if we exclude the case of dependence with no correlation. The two procedures can be combined for uniformly valid, but conservative inference. For pivotal statistics, either procedure also provides pointwise asymptotic refinements over the Gaussian approximation when the limiting distribution is normal. We discuss several special cases and extensions, including V-statistics, subgraph densitities for network data, and non-exhaustive samples of matched data.\\[4pt]
\end{abstract}

\maketitle


\section{Introduction}



We consider a random array $\left(Y_{it}\right)_{i,t}$ where we assume that (1) for $i=1,\dots,N$, the rows $\mathbf{Y}_{i\cdot}:=(Y_{i1},Y_{i2},\dots)$  are assumed to be \emph{conditionally} i.i.d. given the column-wise marginal distributions $F_t(y):=P(Y_{it}\leq y|t)$, $t=1,2,\dots$. (2) For $t=1,\dots,T$ the columns $\mathbf{Y}_{\cdot t}:=(Y_{1t},Y_{2t},\dots)$ are \emph{conditionally} i.i.d. given the row-wise marginal distributions $F_i(y):=P(Y_{it}\leq y|i)$, $i=1,2,\dots$. Otherwise dependence within rows and columns is left unrestricted. In the benchmark case, we observe the realization $Y_{it}$ for each tuple $(i,t)$ where $i=1,\dots,N$ and $t=1,\dots,T$. We later consider extensions to arrays indexed by more than two dimensions as well as to the case in which $Y_{it}$ is observed for only a subset of tuples $(i,t)$.



Our main results concern the problem of bootstrapping the distribution of the sample average
\[\bar{Y}_{NT}:=\frac1{NT}\sum_{i=1}^N\sum_{t=1}^T Y_{it}\]
The bootstrap procedure we propose in this paper is adaptive to features of the joint distribution of the random array, and approximations are as $N$ and $T$ grow large at the same rate.

The leading case of bootstrapping the sample average already reflects the main new technical challenges arising from multi-way clustering. However, we also consider a number of practically relevant extensions and generalizations. For one, the procedure can be easily adapted for statistics that are asymptotically linear (i.e. that can be approximated via influence functions), or differentiable functions of $\bar{Y}_{NT}$. It is also straightforward to implement the procedure to settings with clustering long more than two dimensions, or $D$-adic data where the random array corresponds to group-level outcomes for any subset of $D$ out of the full set of $N$ units included in the sample. Another practically impportant extension concerns the case in which the variable $Y_{it}$ is only observed for a subset of the pairs $\{(i,t):i=1,\dots,N,t=1,\dots,T\}$ (non-exhaustively matched samples).

Arrays with multi-way clustering may result from sampling from an infinite population of ``cross-sectional" and ``temporal" units, where we draw $N$ ``cross-sectional" units $i=1,\dots,N$ and $T$ ``temporal" units $t=1,\dots,T$ independently at random, and independently from one another.

\begin{ex}
\textbf{Static panels.} One interpretation of this setup is a panel in which cross-sectional units are observed over time, and the outcome of interest is subject to both common aggregate shocks and unit-level heterogeneity. Dependence structures of this type are a distinguishing feature of classical differences-in-differences designs that aim to control for average effects of shocks and unobserved heterogeneity. Our framework does not restrict the ``number" of these shocks, or how they may interact in a generative model for the outcome variable $Y_{it}$.
\end{ex}

\begin{ex}
\textbf{Matched data.} Some data sets take the form of matched samples between different groups of units $i=1,\dots,N$ and $t=1,\dots,T$, respectively, where $Y_{it}$ measures an outcome at the level of the match. This setup includes test scores for a random sample of students and teachers, or wages (marginal product of labor) for a random sample of workers and firms. If these units are regarded as random draws from their respective super-populations, the procedure developed in this paper can be used for inference with respect to the distribution of $Y_{it}$ in that super-population. For matched data we often only observe $Y_{it}$ for a small subset of the possible dyads $(i,t)$ (non-exhaustively matched samples), and we discuss an adaptation of our bootstrap method to this case in section \ref{sec:extensions_sec}.
\end{ex}

\begin{ex}\textbf{V- and U-statistics} We can view U-statistics (see e.g. \cite{Vdv98} for definitions and an overview of classical properties) as a special case of our framework for $D$-adic data. For an i.i.d. random sample $X_1,\dots,X_N$, a V-statistic of degree $D$ with a symmetric kernel $h(x_1,\dots,x_D)$ is defined as
\[V = \frac1{N^D}\sum_{i_1\dots i_D}h(X_{i_1},\dots,X_{i_D})\]
which is equal to the $D$-fold sample average $\bar{Y}_{N,D}:=\frac1{N^D}\sum_{i_1\dots i_D}Y_{i_1\dots i_D}$ for the observations
\[Y_{i_1\dots i_D}:=h(X_{i_1},\dots,X_{i_D})\]
The kernel $h(\cdot)$ is called degenerate if $\mathbb{E}[h(x,X_2,\dots,X_D)]$ is constant. The asymptotic behavior of $\bar{Y}_{N,D}$ depends crucially on whether the kernel is degenerate given the distribution of $X_i$. The corresponding U-statistic is
\[U = \binom{N}{D}^{-1}\sum_{i_1<i_2\dots<i_D}h(X_{i_1},\dots,X_{i_D})=\binom{N}{D}^{-1}
\sum_{i_1\dots i_D}w_{i_1\dots i_D}h(X_{i_1},\dots,X_{i_D})\]
where $w_{i_1\dots i_D}=\dum\{i_1<i_2\dots<i_D\}$. Hence U-statistics can be viewed as a special case of a mean for a non-exhaustively matched sample, which is discussed in Section \ref{sec:extensions_sec}.
\end{ex}

\begin{ex}
\textbf{Network data.} The general framework can be applied to subgraph counts or graph/homomorphism densities in networks. Suppose that for a network with $N$ nodes we observe the $N\times N$ adjacency matrix $\mathbf{G}_N$ with entries $G_{ij}$ corresponding to indicators whether that network includes a directed edge from $i$ to $j$, where we assume $G_{ii}=0$ for all $i$ (no self-links). Following the approach in \cite{Lov12}, \cite{BCL11}, and \cite{BBi15}, we can regard $\mathbf{G}_N$ as a sample from an unlabeled infinite graph. For example to evaluate the extent of clustering/triadic closure in the network, we can consider triad-level subgraph counts $T_r:=\frac{6}{N(N-1)(N-2)}\sum_{i<j<k}Y_{ijk,r}$ for $r=2,3$ where $Y_{ijk,2}=G_{ij}G_{ik}$ and $Y_{ijk,3}=G_{ij}G_{ik}G_{jk}$, so that $Y_{ijk,3}=0$ whenever $i,j,k$ are not distinct, and $Y_{ijk,2}=0$ if $i=j$ or $i=k$. With degree heterogeneity across nodes, entries $Y_{ijk,r}$ exhibit dependence across each dimension of the array. Our framework nests problems of this type where subgraphs involving $D$ nodes can be represented in terms of an $D$-dimensional array with a shared index set $\{1,\dots,N\}$ for each dimension.
\end{ex}

Other prominent applications allowing for (generally non-additive) dependence across several dimensions from e-commerce, biogenetics, and crop science are cited in \cite{Owe07}.

Generally speaking, we need to distinguish three scenarios regarding the large-sample distribution of the mean: in the absence of clustering, elements of the array $(Y_{it})$ are mutually independent. When elements are correlated within clusters, the convergence rate of the mean is determined by the number of clusters. Finally, in non-separable models of heterogeneity, elements within a cluster may be dependent even if they are uncorrelated. In that last case, which is specific to clustering in two or more dimensions, the asymptotic behavior of the sample mean is generally non-standard, and the conventional estimator of its asymptotic variance is not consistent. To frame ideas, we next give two stylized examples to illustrate the difference between these three cases.


\begin{ex}\label{ex:add_factor_model}\textbf{Additive Factor Model.} To shape ideas, consider first the case where clustering results from an additive model with cluster-level effects
\[Y_{it} = \mu + \alpha_i+ \gamma_t + \varepsilon_{it}\]
where $\mu$ is fixed and $\alpha_i,\gamma_t,\varepsilon_{it}$ are zero-mean, i.i.d. random variables for $i=1,\dots,N$ and $t=1,\dots,T$ with bounded second moments, and $N=T$. From a standard central limit theorem we find that in the non-degenerate case with $\var(\alpha_i)>0$ or $\var(\gamma_t)>0$, the sample distribution \[\sqrt{N}(\bar{Y}_{NT}-\mathbb{E}[Y_{it}])\rightsquigarrow N(0,\var(\alpha_i)+\var(\gamma_t)),\] whereas in the degenerate case of no clustering, $\var(\alpha_i)=\var(\gamma_t)=0$, \[\sqrt{NT}(\bar{Y}_{NT}-\mathbb{E}[Y_{it}])\rightsquigarrow N(0,\var(\varepsilon_{it}))\]
where ``$\rightsquigarrow$" denotes convergence in distribution.

If the marginal distributions of these three factors were known, we could simulate from the joint distribution of $\left(Y_{it}\right)_{i=1,\dots,N\\t=1,\dots,T}$ by sampling the individual components at random. A bootstrap procedure would replace these unknown distributions with consistent estimates. If the distribution of $\alpha_i$ is not known, an intuitively appealing estimator of $\alpha_i$ is
\[\hat{\alpha}_i:=\frac1T\sum_{t=1}^T(Y_{it} - \bar{Y}_{NT}) = \alpha_i + \frac1T\sum_{t=1}^T(\varepsilon_{it} - \bar{\varepsilon}_{NT})=\alpha_i + O_p(1/\sqrt{T})\] Similarly, we can estimate
$\hat{\gamma}_t:=\frac1N\sum_{i=1}^N(Y_{it}-\bar{Y}_{NT})=\gamma_t + O_P(1/\sqrt{N})$, and $\hat{\varepsilon}_{it}:=Y_{it} - \bar{Y}_{NT} - \hat{\alpha}_i-\hat{\gamma}_t=\varepsilon_{it} + o_p(1)$. Given these estimates, we can form the bootstrap sample $Y_{it}^*:=\bar{Y}_{NT} + \alpha_i^* + \gamma_t^* + \varepsilon_{it}^*$ by drawing with replacement from the estimated marginal distributions of $\alpha_i,\gamma_t,\varepsilon_{it}$, and obtain the bootstrapped mean $\bar{Y}_{NT}^*:=\frac1{NT}\sum_{i=1}^N\sum_{t=1}^T Y_{it}^*$. From simple variance calculations, we find that $\frac{N}{N-1}\var\left(\hat{\alpha}_i-\frac1N\sum_{j=1}^N\hat{\alpha}_j\right)=\var(\alpha_i) + \var(\varepsilon_{it})/T$ and $\frac{T}{T-1}\var\left(\hat{\gamma}_t-\frac1T\sum_{s=1}^T\hat{\gamma}_s\right)=\var(\gamma_t) + \var(\varepsilon_{it})/N$.

Hence, in the non-degenerate case with $\var(\alpha_i)>0$ or $\var(\gamma_t)>0$, the bootstrap distribution \[\sqrt{N}(\bar{Y}_{NT}^*-\bar{Y}_{NT})\rightsquigarrow N(0,\var(\alpha_i)+\var(\gamma_t))\] converges to the same limit as the sampling distribution, so that estimation error in $\hat{\alpha}_i$ does not affect the asymptotic variance. However, in the degenerate case of no clustering, $\var(\alpha_i)=\var(\gamma_t)=0$, the bootstrap distribution \[\sqrt{NT}(\bar{Y}_{NT}^*-\bar{Y}_{NT})\rightsquigarrow N(0,3\var(\varepsilon_{it}))\] asymptotically over-estimates the variance of the sampling distribution, so that this naive bootstrap procedure is inconsistent in the degenerate case.\footnote{Adaptations of the nonparametric bootstrap combining i.i.d. draws of columns and rows of the array $\left(Y_{it}\right)_{i=1,\dots,N\\t=1,\dots,T}$ have been found to have similar problems, see \cite{McC00} and \cite{Owe07}.}
\end{ex}

Our formal representation result below establishes an approximate representation of that form for general case, so we can build on intuitions from this simple example. If we furthermore treat $\varepsilon_{it}$ as a projection error, its distribution may generally depend on $\alpha_i,\gamma_t$.

As the next example illustrates, the non-separable case has added complications from the fact that $\alpha_i,\gamma_t$ may interact. However, in either case the potential complications with the bootstrap stem entirely from the degenerate case.


\begin{ex}\textbf{Non-Gaussian Limit Distribution.} One important insight from the literature on U-statistics is that the limiting behavior of $\bar{Y}_{NT}$ depends crucially on the degree of degeneracy of the row/column projection $U_{NT}$. Specifically, there are forms of dependence in $Y_{it}$ such that the sample mean $\bar{Y}_{NT}$ is not asymptotically normal, even if standard regularity conditions are satisfied by the conditional distributions of $Y_{it}$ given $\alpha_i$, and the conditional distribution given $\gamma_t$, respectively (i.e. in each row and column of the array):

To illustrate the difficulty, we can consider the following example, adapting a classical counterexample for degenerate U-statistics (see e.g. \cite{Bre83})
\[Y_{it} = \alpha_i\gamma_t + \varepsilon_{it}\]
where $\alpha_i,\gamma_t,\varepsilon_{it}$ are independently distributed, with $\mathbb{E}[\varepsilon_{it}]=0$, $\var(\alpha_i)=\sigma_{\alpha}^2$, $\var(\gamma_t)=\sigma_{\gamma}^2$, and $\var(\varepsilon_{it})=\sigma_{\varepsilon}^2$.

If in addition, $\mathbb{E}[\alpha_i]=\mathbb{E}[\gamma_t]=0$, we can use a standard CLT and the continuous mapping theorem to verify that
\begin{eqnarray}\nonumber\sqrt{NT}\bar{Y}_{NT}&=&\frac1{\sqrt{NT}}\sum_{i=1}^N\sum_{t=1}^T(\alpha_i\gamma_t + \varepsilon_{it})\\ \nonumber&=&\left(\frac1{\sqrt{N}}\sum_{i=1}^N\alpha_i\right)\left(\frac1{\sqrt{T}}\sum_{t=1}^T\gamma_t\right) + \frac1{\sqrt{NT}}\sum_{i=1}^N\sum_{t=1}^T\varepsilon_{it}\\
\nonumber&\rightsquigarrow& \sigma_{\alpha}\sigma_{\gamma}Z_1 Z_2 + \sigma_{\varepsilon}Z_3
\end{eqnarray}
where $Z_1,Z_2,Z_3$ are independent standard normal random variables. Since the product of two independent normal random variables is not normally distributed,  $\sqrt{NT}\bar{Y}_{NT}$ is not asymptotically normal.\footnote{Since $Z_1Z_2 = \frac14(Z_1 + Z_2)^2 - \frac14(Z_1 - Z_2)^2$, where $\cov(Z_1+Z_2,Z_1-Z_2) = \var(Z_1)-\var(Z_2)=0$. Hence, $Z_1Z_2 = \frac12(W_1 - W_2)$, where $W_1,W_2$ are independent chi-square random variables with one degree of freedom.} Note also that if instead $\mathbb{E}[\alpha_i]\neq0$ or $\mathbb{E}[\gamma_t]\neq0$ the statistic remains asymptotically normal at the slower $\sqrt{T}$ ($\sqrt{N}$, respectively) rate.
\end{ex}

The case of non-separable heterogeneity with row- and column-means centered at zero is an example of dependence in the absence of within-cluster correlation in the first moments of $Y_{it}$. For that scenario, plug-in asymptotic inference based on the normal distribution is not valid, and we find that the default estimator for the asymptotic variance is inconsistent due to the within-cluster correlation in second moments of $Y_{it}$. Moreover, we show that uniformly consistent estimation of the limiting distribution is in fact impossible, rather the proposed methods are pointwise consistent and for standard inference problems uniformly valid, but conservative procedures can be easily obtained from these. Interestingly, this case is not relevant for the limiting distribution of the sample mean when observations are clustered in at most one dimension.


\subsection{Contribution and Related Literature}

With clustering in multiple dimensions, the problem of resampling is fundamentally different from the case of independent clusters, since the structure of the data no longer implies finite or weak dependence across units. In fact, \cite{McC00} showed that there exists no straightforward adaptation of the classical nonparametric bootstrap (\cite{Efr79}, see also \cite{Hal92}, and \cite{Hor01} for an exposition) that is consistent with multi-way clustered data.\footnote{\cite{McC00}'s argument goes as follows: there is no consistent estimator for the variance of the sample mean that is a nonnegative quadratic function of the observations $Y_{it}$. In particular the bootstrapped variance from any resampling scheme that draws directly from the original values of the variable of interest is a function of this type, and therefore such a bootstrap scheme cannot be consistent. We propose a hybrid scheme that does not fall under his narrower definition of the bootstrap.} Our procedure combines features of the nonparametric bootstrap with those of the wild bootstrap (\cite{Wu86} and \cite{Liu88}) to achieve (pointwise) consistency in each case, as well as uniformity and refinements for cases in which the limiting behavior of the statistic is standard. We find that the problem of multi-way clustering has a natural connection to the theory of U- and V-statistics, separate bootstrap procedures for which have been proposed by \cite{Bre83} and \cite{AGi92} in the non-degenerate and degenerate cases. When applied to V-statistics, our procedure is (pointwise) adaptive when the degree of degeneracy of the kernel is unknown.

Asymptotic standard errors with multi-way clustering have been proposed by \cite{CGM11}, and can be used for ``plug-in" asymptotic inference in the Gaussian limiting case - see also \cite{CMi14} and \cite{ASA15} for the case of dyadic data. The ``pigeonhole" bootstrap proposed by absence of clustering. Subsample bootstraps, including the method by \cite{BBi15} for network data, adapt quite naturally to features of the data-generating process and are particularly attractive when evaluation of the statistic over the full sample is computationally very costly. However, even for well-behaved cases it is not known whether favorable properties are uniform, or whether the procedure achieves refinements over ``plug-in" asymptotics.

Our findings regarding the non-existence of uniformly consistent estimators for this problem - including the case of U-statistics with kernel of unknown order of degeneracy - is new to the literature. The problem can be thought of as a further instance of a discontinuity in the pointwise asymptotic limiting distribution when a relevant parameter is on the boundary of the parameter space (see \cite{And00}, \cite{And01}, and \cite{Agu10}). Our analysis benefits from insights and techniques developed for that more general problem.


\subsection{Notation and Overview} Throughout the paper, we use $\mathbb{P}$ to denote the joint distribution of the array $\left(Y_{it}\right)_{i,t}$, and denote drifting data-generating processes (DGP) indexed by $N,T$ with $\mathbb{P}_{NT}$. The bootstrap distribution for $\left(Y_{it}^*\right)$ given the realizations $(Y_{it}:i=1,\dots,N;t=1,\dots,T)$ is denoted $\mathbb{P}_{NT}^*$. We denote expected values under these respective distributions using $\mathbb{E},\mathbb{E}_{NT}$, and $\mathbb{E}_{NT}^*$, respectively.

In the remainder of the paper, we first establish a representation for the array $(Y_{it})$ which is then used to motivate a bootstrap procedure. Formal results regarding consistency and refinements for that bootstrap procedure are given in Section 4. We furthermore give several generalizations of the main procedure and illustrate its performance using Monte Carlo simulations.

\section{Representation}
This section develops a stochastic representation for $\bar{Y}_{NT}$ as a function of sample means of uncorrelated factors, where averages can be taken separately in each dimension of the random array. After establishing joint convergence of these more elementary sample averages, we can then develop a limit theory based on this representation.

We first show that the array $(Y_{it})_{i,t}$ permits a decomposition of the form
\[Y_{it} = b + a_i + g_t + w_{it},\hspace{0.5cm}\mathbb{E}[w_{it}|a_i,g_t] = 0\]
where $a_i$ and $g_t$ are mean-zero and mutually independent, so that the joint distribution of $Y_{it}$ can then be described in terms of the respective marginal distributions of $a_i$ and $g_t$, and the conditional distribution of $w_{it}$ given $a_i,g_t$.



Such a representation is immediate for the leading example of the additive factor model in Example \ref{ex:add_factor_model}, and we now show that it is in fact without loss of generality for arrays exhibiting dependence in two or more dimensions. Specifically, since the rows (and columns, respectively) of the array $(Y_{it})_{i,t}$ are i.i.d., Theorem 1.4 in \cite{Ald81} implies that we can write
\[Y_{it}=f(\alpha_i,\gamma_t,\varepsilon_{it})\]
for some function $f(\cdot)$, where $\alpha_1,\dots,\alpha_N$, $\gamma_1,\dots,\gamma_T$ and $\varepsilon_{11},\dots,\varepsilon_{NT}$ are mutually independent, uniformly distributed random variables. This representation is not restricted to the case of two-way dependence, see \cite{Hoo79} for a generalization to partial exchangeability in more than two dimensions.

%
%

If the relevant conditional expectations are well-defined, we can represent $Y_{it}$ via the projection expansion
\begin{eqnarray}\nonumber Y_{it} &=& \mathbb{E}[Y_{it}]+ (\mathbb{E}[Y_{it}|\alpha_i]-\mathbb{E}[Y_{it}]) + (\mathbb{E}[Y_{it}|\gamma_t]-\mathbb{E}[Y_{it}])\\
\nonumber&&+(\mathbb{E}[Y_{it}|\alpha_i,\gamma_t] - \mathbb{E}[Y_{it}|\alpha_i]-\mathbb{E}[Y_{it}|\gamma_t]+\mathbb{E}[Y_{it}]) +(Y_{it}-\mathbb{E}[Y_{it}|\alpha_i,\gamma_t])\\
\label{Y_it_decomp}&=:&b+a_i + g_t + v_{it}+e_{it}\end{eqnarray}
where we define $e_{it}=Y_{it}-\mathbb{E}[Y_{it}|\alpha_i,\gamma_t]$, $a_i:=\mathbb{E}[Y_{i1}|\alpha_i]-\mathbb{E}[Y_{i1}]$,
$g_t=\mathbb{E}[Y_{1t}|\gamma_t]-\mathbb{E}[Y_{1t}]$,
$v_{it}=\mathbb{E}[Y_{it}|\alpha_i,\gamma_t] - \mathbb{E}[Y_{it}|\alpha_i]-\mathbb{E}[Y_{it}|\gamma_t] + \mathbb{E}[Y_{it}]$, and $b=\mathbb{E}[Y_{it}]$. Since temporal and cross-sectional units were drawn independently, $a_1,\dots,a_N$ and $g_1,\dots,g_T$ are independent of each other. Also by construction, $\mathbb{E}[e_{it}|a_i,g_t,v_{it}]=0$ and $\mathbb{E}[v_{it}|a_i,g_t] = 0$. In particular, the terms $e_{it},(a_i,g_t),v_{it}$ are uncorrelated.

Given this representation, we can rewrite the sample mean
\[\bar{Y}_{NT} = b + \bar{a}_N + \bar{g}_T + \bar{v}_{NT} + \bar{e}_{NT}\]
where $\bar{a}_N:=\frac1N\sum_{i=1}^Na_i$, $\bar{g}_T:=\frac1T\sum_{t=1}^T g_t$,  $\bar{v}_{NT}:=\frac1{NT}\sum_{t=1}^T\sum_{i=1}^Nv_{it}$, and $\bar{e}_{NT}:=\frac1{NT}\sum_{t=1}^T\sum_{i=1}^Ne_{it}$. We also denote the unconditional variances of the projections with $\sigma_a^2:=\var(a_i)$, $\sigma_g^2:=\var(g_t)$, $\sigma_v^2:=\var(v_{it})$, and $\sigma_e^2:=\var(e_{it})$, respectively. We also let $w_{it}:=v_{it} + e_{it}$ and denote its variance by $\sigma_w^2 = \var(w_{it})$.


Throughout the remainder of the paper, we are going to maintain the following conditions on the distribution of the random array:

\begin{ass}\label{integ_ass}\textbf{(Integrability)} (a) Let $Y_{it}=f(\alpha_i,\gamma_t,\varepsilon_{it})$ where $(\alpha_i)_{i}$, $(\gamma_t)_{t}$, and $\left(\varepsilon_{it}\right)_{i,t}$ are random arrays whose elements are i.i.d.. (b) The random variables $a_i/\sigma_a$, $g_t/\sigma_g$, $v_{it}/\sigma_v$, and $e_{it}/\sigma_e$ have bounded moments up to the order $4+\delta$ for some $\delta>0$ whenever the respective variances $\sigma_a^2,\sigma_g^2,\sigma_v^2,\sigma_e^2>0$. (c) $\sigma_v^2+\sigma_e^2>0$. 
\end{ass}

For our analysis, it is instructive to interpret the row/column projection of the sample average,
\[\bar{a}_N + \bar{g}_T + \bar{v}_{NT}\equiv\frac1{NT}\sum_{t=1}^T\sum_{i=1}^N\mathbb{E}[Y_{it}|\alpha_i,\gamma_t]-\mathbb{E}[Y_{it}]
=:\frac1{NT}\sum_{t=1}^T\sum_{i=1}^Nh(\alpha_i,\gamma_t)\]
as a generalized (two-sample) U-statistic with a kernel $h(\alpha,\gamma)$ evaluated at the samples $\alpha_1,\dots,\alpha_N$ and $\gamma_1,\dots,\gamma_T$, respectively. In that framework, the terms $\bar{a}_N + \bar{g}_T$ and $\bar{v}_{NT}$ constitute the first and second-degree projection terms in a Hoeffding decomposition of the quantity $U_{NT}=\frac1{NT}\sum_{t=1}^T\sum_{i=1}^N\mathbb{E}[Y_{it}|\alpha_i,\gamma_t]-\mathbb{E}[Y_{it}]$.

The full problem of characterizing the distribution of $\bar{Y}_{NT}$ differs from the classical analysis of U-statistics for one in the presence of the projection error $e_{it}$, and also in that the factors $\alpha_i,\gamma_t$ are not observable data, but implicitly defined by Aldous' (1981) construction. Nevertheless, we show below that these additional aspects do not preclude us from applying on insights and techniques for U-statistics to the present problem.

Specifically, we can show that we can approximate the sample and bootstrap distributions of the statistic by a function of sample averages of independent random variables. Define
\[h(\alpha,\gamma):=\mathbb{E}[Y_{it}|\alpha_i=\alpha,\gamma_t=\gamma]-\mathbb{E}[Y_{it}]\]
Under Assumption \ref{integ_ass}, the integral operator
\[S(u)(g) = \int h(a,g)u(a)F_{\alpha}(da) \]
and its adjoint
\[S^*(u)(a) = \int h(a,g) u(g) F_{\gamma}(dg)\]
are both compact, so that the spectral representation theorem permits the low-rank approximation
\begin{equation}\label{spectral_rep}h(\alpha,\gamma) = \sum_{k=1}^{\infty}c_k\phi_k(\alpha)\psi_k(\gamma)\end{equation}
under the $L_2(F_{\alpha,\gamma})$ norm on the space of smooth functions of $(\alpha,\gamma)\in[0,1]^2$. Here, $\left(c_k\right)_{k\geq1}$ is a sequence of singular values with $\lim|c_k|\rightarrow0$, and $\left(\phi_k(\cdot)\right)_{k\geq1}$ and $\left(\psi_k(\cdot)\right)_{k\geq1}$ are orthonormal bases for $L_2([0,1],F_{\alpha})$ and $L_2([0,1],F_{\gamma})$, respectively.

Given this representation, we can write
\[\frac1{NT}\sum_{i=1}^N\sum_{t=1}^T(h(\alpha_i,\gamma_t)-a_i-g_t) = \sum_{k=1}^{\infty}c_k\left(\frac1N\sum_{i=1}^N(\phi_k(\alpha_i)-\mathbb{E}[\phi_k(\alpha_i)])\right)
\left(\frac1T\sum_{t=1}^T(\psi_k(\gamma_t)-\mathbb{E}[\psi_k(\gamma_t)])\right)\]
so that the second-order projection term can also be represented as a function of countably many sample averages of i.i.d., mean-zero random variables.

We find that point-wise consistency of the bootstrap does not require any additional conditions on the conditional expectation function $h(\alpha,\gamma)$ beyond Assumption \ref{integ_ass}. For the uniform consistency results which include the case in which the asymptotically non-Gaussian component is of first order, we need to restrict the eigenfunctions and coefficients in the spectral representation (\ref{spectral_rep}).

\begin{ass}\label{spectral_ass}
The conditional mean function $h(\alpha,\gamma):=\mathbb{E}[Y_{it}|\alpha_i=\alpha,\gamma_t=\gamma]$ admits a spectral representation
\[h(\alpha,\gamma) = \sum_{k=1}^{\infty}c_k\phi_k(\alpha)\psi_k(\gamma)\]
under the $L_2(F_{\alpha,\gamma})$ norm, where (a) the singular values are uniformly bounded by a null sequence $\bar{c}_k\rightarrow0$, that is $c_k\leq \bar{c}_k$ for each $k=1,2,\dots$, and (b) The first three moments of the eigenfunctions $\phi_k(\alpha_i)$ and $\psi_k(\gamma_t)$ are bounded by a constant $B>0$ for each $k=1,2,\dots$.
\end{ass}

Imposing common bounds on moments and singular values restricts the set of joint distributions $F$ for the array to a uniformity class, where the sequence $(a_k)_{k\geq0}$ controls the magnitude of the error from a finite-dimensional approximation to $h(\alpha,\gamma)$, where we truncate the expansion in (\ref{spectral_rep}) after a finite number of summands $k=1,\dots,K$. Comparable high-level conditions on spectral approximations are commonly used to define uniformity classes in nonparameric estimation of operators, see e.g. \cite{HHo05} and \cite{CFR07}.

\section{Bootstrap Procedure}


\label{sec:algorithm_sec}

The previous discussion shows that the rate of convergence and the limiting distribution of the sample mean $\bar{Y}_{NT}-\mathbb{E}[Y_{it}]$ depend crucially on the different scale parameters introduced above. If observations are independent across rows and columns, then $\sqrt{NT}(\bar{Y}_{NT}-\mathbb{E}[Y_{it}])\stackrel{d}{\rightarrow} N(0,\var(e_{it}))$. If within-cluster covariances are bounded away from zero in at least one dimension, then $\sqrt{N}(\bar{Y}_{NT}-\mathbb{E}[Y_{it}])\stackrel{d}{\rightarrow} N(0,\max\{\var(a_i),\var(g_t)\})$. Our aim is to obtain a bootstrap procedure that is adaptive and interpolates between the degenerate and non-degenerate cases. We consider pointwise and uniform consistency of the bootstrap as well as (pointwise) refinements, and find that these properties generally hold under different subsets of conditions on the data generating process.

In particular, for the performance of the bootstrap it is crucial at what rate(s) estimators for the different model components are consistent depending on the extent of clustering in the true DGP. Most importantly, the variance of the projection terms $\hat{a}_i$ and $\hat{g}_t$ is $\sigma_a^2+\sigma_w^2/T$ and $\sigma_g^2+\sigma_w^2/N$, respectively, so that the ``convolution" components depending on $\sigma_w^2$ dominates in the degenerate case. In order to correct for the contribution of the row/column averages of $w_{it}$ we would therefore want to shrink the scale of the distribution of $\hat{a}_i,\hat{g}_t$ by the variance ratio
\[\lambda = \frac{T\sigma_a^2 + N\sigma_g^2}{T\sigma_a^2 + N\sigma_g^2+2\sigma_w^2}\]
In the bootstrap procedure we replace the unknown variances with unbiased and consistent estimators
\begin{eqnarray}\nonumber \hat{\sigma}_w^2&:=&\frac1{NT-N-T}\sum_{i=1}^N\sum_{t=1}^T(Y_{it}-\bar{Y}_{iT}-\bar{Y}_{Nt}+\bar{Y}_{NT})^2,\\ \nonumber\hat{\sigma}_a^2&:=&\frac1{N-1}\sum_{i=1}^n(\bar{Y}_{iT}-\bar{Y}_{NT})^2 - \frac1T\hat{\sigma}_w^2,\textnormal{ and}\\ \nonumber\hat{\sigma}_g^2&:=&\frac1{T-1}\sum_{t=1}^T(\bar{Y}_{Nt}-\bar{Y}_{NT})^2 - \frac1N\hat{\sigma}_w^2\end{eqnarray}
to obtain alternative estimators for $\lambda$. Specifically, we let
\[\hat{\lambda}:=\frac{\max\left\{0,T\hat{\sigma}_a^2 + N\hat{\sigma}_g^2\right\}}{T\hat{\sigma}_a^2 + N\hat{\sigma}_g^2+\hat{\sigma}_w^2}\]
where we take the maximum with zero in the numerator to ensure that $\hat{\lambda}$ is nonnegative in finite sample. We find that $\hat{\lambda}$ is uniformly consistent for $\lambda$ if and only if $\bar{v}_{NT}$ does not contribute to the limiting distribution. In order to obtain (pointwise) consistency in all cases, we employ the alternative estimator
\[\tilde{\lambda}:=\hat{\lambda}\dum\left\{\hat{\sigma}_a^2\geq \frac{\kappa_T}T\textnormal{ or }\hat{\sigma}_g^2\geq\frac{\kappa_N}N\right\}\]
for some slowly increasing $\kappa_N,\kappa_T$ (e.g. $\kappa_N = \log N$ and $\kappa_T = \log T$).\footnote{Pointwise consistent model selection when a parameter relevant for the asymptotic distribution is near or at the boundary of the parameter space was first considered for the bootstrap by \cite{And00}.} We also show that allowing for the case in which $\bar{v}_{NT}$ constributes to the limiting distribution, uniformly consistent estimation of the limiting distribution is not possible, neither using the bootstrap nor any alternative method.

For the leading case of exhaustive sampling with clustering in two dimensions, we then propose the following resampling algorithm to estimate the sampling distribution:

\begin{itemize}
\item[(a)] Use the sample to obtain $\hat{a}_i:=\frac1T\sum_{t=1}^T(Y_{it}-\bar{Y}_{NT})$, $\hat{g}_t:=\frac1N\sum_{i=1}^N(Y_{it}-\bar{Y}_{NT})$, and $\hat{w}_{it}:=Y_{it} - \hat{a}_i - \hat{g}_t + \bar{Y}_{NT}$.
\item[(b)] For the $b$th bootstrap iteration, draw $a_{i,b}^*:=\hat{a}_{k_b^*(i)}$ and $g_{t,b}^*:=\hat{g}_{s_b^*(t)}$, where $k_b^*(i)$ and $s_b^*(t)$ are i.i.d. draws from the discrete uniform distribution on the index sets $\{1,\dots,N\}$ and $\{1,\dots,T\}$, respectively.
\item[(c)] Generate $w_{it,b}^*:=\omega_{1i,b}\omega_{2t,b} \hat{w}_{k_b^*(i)s_b^*(t)}$, where $\omega_{1i,b},\omega_{2t,b}$ are i.i.d. random variables with $\mathbb{E}[\omega_{\cdot}]=0,\mathbb{E}[\omega_{\cdot}^2]=\mathbb{E}[\omega_{\cdot}^3]=1$
\item[(d)] For a given choice $\hat{\lambda}$ of an estimator for $\lambda$, generate a bootstrap samples of draws $Y_{it,b}^* = \bar{Y}_{NT} +  \sqrt{\hat{\lambda}}(a_{i,b}^* + g_{t,b}^*) + w_{it,b}^*$  and obtain the bootstrapped statistic
$\bar{Y}_{NT,b}^*:=\frac1{NT}\sum_{i=1}^N\sum_{t=1}^T Y_{it,b}^*$.
\item[(e)] Repeat $B$ times and approximate the distribution of $r_{NT}(\bar{Y}_{NT}^*-\bar{Y}_{NT})$ using the empirical distribution over the bootstrap draws $\bar{Y}_{NT,1}^*,\dots,\bar{Y}_{NT,B}^*$.
\end{itemize}


We discuss a choice the distribution of the random variable $\omega_{\cdot}$ in the appendix, where we adjust the popular two-point specification proposed by \cite{Mam92} to correct for finite-sample bias in the second and third moments of the empirical distribution. Such a correction is not needed for the theoretical properties of our procedure but turns out to yield some improvements in simulations for small values of $N,T$.

\section{Theoretical Properties}


The limiting behavior of the sample mean $\bar{Y}_{NT}-\mathbb{E}[Y_{it}]$ is in part determined by the variances of the components of the decomposition in (\ref{Y_it_decomp}) with
$\sigma_a^2 = \var(a_i)$, $\sigma_g^2 = \var(g_t)$, $\sigma_v^2 = \var(v_{it})$, $\sigma_e^2 = \var(e_{it})$, and $\sigma_w^2 = \var(w_{it}) = \sigma_v^2 + \sigma_e^2$,
where we may also consider drifting sequences of distributions, where these parameters change as $N$ and $T$ grow to infinity. We also let
\[r_{NT}^2:=N^{-1}\sigma_a^2 + T^{-1}\sigma_g^2 + (NT)^{-1}\sigma_w^2\]
We maintain throughout that either $\sigma_g^2 + \sigma_a^2>0$ or $\sigma_w^2>0$, and that $N$ and $T$ grow at the same rate as we take limits.

\subsection{Bootstrap Consistency}

In order to establish uniform convergence with respect to the joint distribution of $(Y_{it})_{i,t}$, we need to consider limits along any drifting sequences for the parameters $\sigma_a,\sigma_g,\sigma_e,\sigma_v$. We then parameterize the limiting distribution with the respective limits of the normalized sequences,
\begin{eqnarray}\nonumber q_{a,NT}:= r_{NT}^{-2}N^{-1}\sigma_a^2,&\hspace{0.3cm}&q_{g,NT}:= r_{NT}^{-2}T^{-1}\sigma_g^2\\
q_{e,NT}:= r_{NT}^{-2}(NT)^{-1}\sigma_e^2&&q_{v,NT}:= r_{NT}^{-2}(NT)^{-1}\sigma_v^2
\end{eqnarray}
We also let $\mathbf{q}_{NT}:=(q_{a,NT},q_{g,NT},q_{e,NT},q_{v,NT})$. From the definition of $r_{NT}$, it follows that the local parameters $q_{a,NT},q_{g,NT},q_{e,NT},q_{v,NT}\in[0,1]$ and $q_{a,NT}+q_{g,NT}+q_{e,NT}+q_{v,N}=1$.


Given the local parameter $\mathbf{q}:=(q_a,q_g,q_e,q_v)$ and the sequence of coefficients in the expansion of $h(\alpha,\gamma)$, $\mathbf{c}:=(c_1,c_2,\dots)$, we define the law
\[\mathcal{L}_0(\mathbf{q},\mathbf{c}):=\sqrt{q_a + q_g + q_e}Z_0 + \sqrt{q_v}V\]
along each converging sequence, where $V:=\lim_{N,T}\frac1{\sigma_v}\sum_{k=1}^{\infty}c_kZ_k^{\psi}Z_k^{\phi}$ with the coefficients $c_k$ potentially variying along the limiting sequence, and $Z_0,Z_1^{\phi},Z_1^{\psi},Z_2^{\phi},Z_2^{\psi},\dots$ are i.i.d. standard normal random variables.

We first give the limit for the sampling distribution of $\bar{Y}_{NT}$:

\begin{thm}\textbf{(CLT for Sampling Distribution)}\label{sample_clt_thm} Suppose that Assumption \ref{integ_ass} holds. Then (a) along any convergent sequence $\mathbf{q}_{NT}\rightarrow\mathbf{q}$, we have
\[\|\mathbb{P}_{NT}(r_{NT}(\bar{Y}_{NT}-\mathbb{E}[Y_{it}]))-\mathcal{L}_0(\mathbf{q},\mathbf{c})\|_{\infty}
\rightarrow0\]
where $\|\cdot\|_{\infty}$ denotes the Kolmogorov metric and $\mathbf{c}=(c_1,c_2,\dots)$. (b) If in addition Assumption \ref{spectral_ass} holds, then the conclusion of (a) also holds under drifting sequences $\mathbf{c}_{NT}\rightarrow\mathbf{c}$.
\end{thm}

See the appendix for a proof. Note that convergence in part (a) is point-wise with respect to the conditional mean function $\frac1{q_{v,NT}}\mathbb{E}[Y_{it}|\alpha_i=\alpha,\gamma_t=\gamma]$, whereas part (b) gives  uniform convergence within the class of distributions satisfying Assumption \ref{spectral_ass}. Using similar arguments, we also obtain the limit of the bootstrap distribution. The following bootstrap CLT is also proven in the appendix:

\begin{thm}\textbf{(Bootstrap CLT)}\label{bootstrap_clt_thm} Suppose that Assumption \ref{integ_ass} holds. Then (a) along any convergent sequence $\mathbf{q}_{NT}\rightarrow\mathbf{q}$ we have
\[\|\mathbb{P}^*_{NT}(r_{NT}(\bar{Y}_{NT}^*-\bar{Y}_{NT}))-\mathcal{L}_0(\mathbf{q},\mathbf{c})\|_{\infty}
\stackrel{p}{\rightarrow}0\]
where $\mathbf{c}=(c_1,c_2,\dots)$, provided that the estimator for $\lambda$ is consistent along that sequence. (b) If in addition Assumption \ref{spectral_ass} holds, then the conclusion of (a) also holds under drifting sequences $\mathbf{c}_{NT}\rightarrow\mathbf{c}$.
\end{thm}

Theorem \ref{bootstrap_clt_thm} indicates that the asymptotic properties of the bootstrap depend crucially on our ability to estimate the variances of the individual projection components at respective rates that are fast enough to ensure convergence of $\hat{\lambda}$ to $\lambda$. Hence, as an intermediate step we establish rates of consistency for the estimators for the respective variances of the projection components, $\hat{\sigma}_a^2,\hat{\sigma}_g^2,\hat{\sigma}_w^2$ introduced in section \ref{sec:algorithm_sec}.

\begin{lem}\label{var_comp_rate_lem} Suppose Assumption \ref{integ_ass} holds. Then (a)
\begin{eqnarray}
\nonumber\hat{\sigma}_a^2 -\sigma_a^2&=&
O_P\left(N^{-1/2}\left(\sigma_a+T^{-1/2}\sigma_e\right)^2+T^{-1}\sigma_v^2\right)\\
\nonumber\hat{\sigma}_g^2-\sigma_g^2&=&O_P\left(T^{-1/2}\left(\sigma_g+ N^{-1/2}\sigma_e\right)^2+N^{-1}\sigma_v^2\right)\\
\nonumber\hat{\sigma}_w^2 - \sigma_w^2 &=& O_P\left((NT)^{-1/2}\sigma_e^2 + (N^{-1/2} + T^{-1/2})\sigma_v^2\right)
\end{eqnarray}
(b) There exist no estimators for $\sigma_a^2,\sigma_g^2$ and $\sigma_w^2$ that converge at rates faster than those given in (a). Specifically, $\sigma_a^2$ cannot be estimated at a rate faster than $T^{-1}$ even when $\sigma_a^2=0$.
\end{lem}

See the appendix for a proof. The lemma implies in particular that the estimators $\hat{\sigma}_a^2,\hat{\sigma}_g^2$ and $\hat{\sigma}_w^2$ are rate-optimal. Together with the continuous mapping theorem, this Lemma implies directly that $\tilde{\lambda}$ is pointwise consistent, and $\hat{\lambda}_{NT}$ is uniformly consistent only if $q_v=0$.


\begin{rem}\textbf{(Estimability of Asymptotic Distribution)}\label{impossibility_rem} Part (b) of Lemma \ref{var_comp_rate_lem} implies that along sequences $\sigma_{a}^2,\sigma_{g}^2,\sigma_{w}^2$ with $\lim_TT\sigma_a^2=q_a$, $\lim_N\sigma_g^2=q_g$ and $\lim_{N,T}\sigma_w^2=q_v+q_e$, the asymptotic variance of the sample mean \[\lim_{N,T}\sqrt{NT}(\bar{Y}_{NT}-\mathbb{E}[Y_{it}])=\lim_{N,T} \left(T\sigma_a^2 + N\sigma_g^2 + \sigma_w^2\right)=q_a + q_g + q_v + q_e\]
cannot be estimated consistently unless $q_v=0$ or $q_a=q_g=0$. If the asymptotic variance cannot be estimated consistently along a particular parameter sequence, it follows in particular that the asymptotic distribution of $\bar{Y}_{NT}$ cannot be estimated consistently uniformly over the entire parameter space, using the bootstrap or any other method.

For a pointwise consistent estimator of the asymptotic variance, let $\kappa_N,\kappa_T$ be sequences of nonnegative numbers that are increasing at a rate slower than $N$ and $T$, respectively, and
\[\hat{S}_{NT}^2:=\left\{\begin{array}{lcl}T\hat{\sigma}_a^2 + N\hat{\sigma}_g^2 + \hat{\sigma}_w^2&\hspace{0.3cm}&
\textnormal{if }\hat{\sigma}_a^2>\kappa_T/T\textnormal{ or }\hat{\sigma}_g^2>\kappa_N/N\\
\hat{\sigma}_w^2&&\textnormal{otherwise}\end{array}\right.\]
Given Lemma \ref{var_comp_rate_lem} (a), it is then possible to verify that indeed \[\left|\var\left(r_{NT}(\bar{Y}_{NT}-\mathbb{E}[Y_{it}])\right) - r_{NT}^{-2}NT\hat{S}_{NT}^2\right|\stackrel{p}{\rightarrow}0\] pointwise for any values of $\sigma_a^2,\sigma_g^2,\sigma_v^2,\sigma_e^2$.
\end{rem}

Similarly, we can use Lemma \ref{var_comp_rate_lem} to establish pointwise consistency of $\tilde{\lambda}$ for $\lambda$ and uniform consistency of $\hat{\lambda}$ if we exclude the case $q_v>0$. Combining this with the sample and bootstrap CLTs to obtain the following consistency result for the bootstrap:

\begin{thm}\textbf{(Bootstrap Consistency)}\label{bootstrap_cons_thm} Suppose that Assumption \ref{integ_ass} holds. Then (a) the sampling distribution $\mathbb{P}(r_{NT}(\bar{Y}_{NT}-\mathbb{E}[Y_{it}]))$ and the bootstrap distribution $\mathbb{P}^*(r_{NT}(\bar{Y}_{NT}-\mathbb{E}[Y_{it}]))$ using $\tilde{\lambda}$ as an estimator for $\lambda$ converge in probability to the same limit,
\[\|\mathbb{P}_{NT}^*(r_{NT}(\bar{Y}_{NT}^*-\bar{Y}_{NT}))-\mathbb{P}_{NT}(r_{NT}(\bar{Y}_{NT}-\mathbb{E}[Y_{it}]))\|_{\infty}
\stackrel{p}{\rightarrow}0\]
(b) If we use the estimator $\hat{\lambda}$ for $\lambda$ and furthermore Assumption \ref{spectral_ass} holds, then convergence is uniform if and only if $q_v=0$.
\end{thm}

See the appendix for a proof.

\begin{rem} Note that these results also applies to generalized (two-sample) U-statistics, which constitute a special case of our setup with $\sigma_e^2=0$. Specifically, the impossibility result in Remark \ref{impossibility_rem} implies that if the order of degeneracy of the kernel is unknown, it is not possible to estimate the distribution of a U-statistic uniformly consistently. The bootstrap procedure in this paper is pointwise adaptive with respect to the order of degeneracy of the kernel of the U-statistic. For classical U-statistics with kernel function of order $D$, we can obtain the analogous result using an adaptation of our bootstrap procedure to $D$-adic data, see Section \ref{sec:extensions_sec} below for a discussion.
\end{rem}

\begin{rem} For certain inference problems it is possible to obtain uniformly valid inference based on a conservative initial estimate $\bar{\lambda}$ for the ratio $\lambda:=(q_a+q_g)/(q_a+q_g+q_e+q_v)$. Specifically, if $\hat{\sigma}_a^2\geq \frac{\kappa_T}T$ or $\hat{\sigma}_g^2\geq\frac{\kappa_N}N$ for some slowly increasing sequences $\kappa_N,\kappa_T$ (e.g. $\kappa_N = \log N$ and $\kappa_T = \log T$), let $\bar{\lambda}=\hat{\lambda}$ and implement the bootstrap algorithm as before. If on the other hand $\hat{\sigma}_a^2< \frac{\kappa_T}T$ and $\hat{\sigma}_g^2<\frac{\kappa_N}N$, we can choose $\bar{\lambda}_{NT}=\frac{\kappa_T+\kappa_N}{\kappa_T + \kappa_N + \hat{\sigma}_w^2}$. Note that since the projection components are mean-zero and uncorrelated, the bootstrap distribution is increasing in the shrinkage parameter $\lambda$ with respect to second-order stochastic dominance (SOSD). Hence using a conservative upper bound for $\lambda$ results in an estimator that SOSD-dominates the sampling distribution with probability approaching 1.
\end{rem}

\subsection{Refinements}

We next consider refinements in the approximation to the distribution of the studentized mean. Specifically, consider the estimator of the asymptotic variance of the sample mean, $\hat{S}_{NT}:=\left(T\hat{\sigma}_a^2 + N\hat{\sigma}_g^2 + \hat{\sigma}_w^2\right)^{1/2}$ and its bootstrap analog $\hat{S}_{NT}^*$, and the studentized sample means $r_{NT}(\bar{Y}_{NT}-\mathbb{E}[Y_{it}])/\hat{S}_{NT}$ and its bootstrap analog $r_{NT}(\bar{Y}_{NT}^*-\bar{Y}_{NT})/\hat{S}_{NT}^*$. We find that the bootstrap approximation provides pointwise refinements for the case in which the limiting distribution for the studentized mean is Gaussian. However, it is important to note that refinements can in general not be obtained for certain special cases.

For one, if the ``Wiener chaos" term remains relevant in the limiting distribution  $\mathcal{L}(\mathbf{q},\mathbf{c})$, i.e. for $q_v>0$, the statistic is no longer asymptotically pivotal. Rather it generally depends on relative weights of the Gaussian component $Z$, and the Wiener chaos component $V$. Hence we cannot expect the bootstrap to provide refinements for this case.

Furthermore, elementary moment calculations reveal that \[\mathbb{E}[\hat{a}_i^3]=\mathbb{E}[a_i^3] + \frac2T\mathbb{E}[a_ie_{it}^2] + \frac1{T^2}\mathbb{E}[e_{it}^3]\]
where the cross-term $\mathbb{E}[a_ie_{it}^2]$ is generally non-zero unless $\mathbb{E}[e_{it}^2|a_i]=0$.
Hence under drifting sequences for the second and third moments of $a_i$, the first term on the right-hand side of that expression need not dominate in the limit, in which case the bootstrap distribution does not match the third moment of $a_i$ under the sampling distribution. Hence, we can in general not obtain a refinement along drifting sequences even when $q_v=0$.

Hence we restrict our attention to pointwise refinements for the case of a Gaussian limiting distribution and can now state the following result:

\begin{thm}\textbf{(Refinements)}\label{refinement_thm} Suppose that Assumption \ref{integ_ass} holds with $\delta>2$. Then, if $\sigma_a^2+\sigma_g^2>0$ or $\sigma_v^2=0$ we have
\[\left\|\mathbb{P}_{NT}^*\left(r_{NT}\left(\frac{\bar{Y}_{NT}^*-\bar{Y}_{NT}}{\hat{S}_{NT}^*}\right)\right)
-\mathbb{P}_{NT}\left(r_{NT}\left(\frac{\bar{Y}_{NT}-\mathbb{E}[Y_{it}]}{\hat{S}_{NT}}\right)\right)\right\|_{\infty}
=O_P(r_{NT}^{-1})\]
point-wise in the distribution of the array $(Y_{it})_{i=1,\dots,N\\t=1,\dots,T}$.
\end{thm}

See the appendix for a proof. Our argument uses \cite{Mam92}'s result based on moment expansions of the statistic rather than the more classical approach based on Edgeworth expansions (see e.g. \cite{Liu88}). This allows us to include the case of a lattice distribution for the random variables $\omega_{1i},\omega_{2t}$ in the implementation of the Wild bootstrap, including the two-point distribution described before.

\section{Extensions}

\label{sec:extensions_sec}
This section gives various extensions to the baseline case. We first consider clustering across $D$ rather than two dimensions, then problems in which data concerns outcomes at the level of a dyad or larger subgroup out of a sample of $N$ ``fundamental" units. Sample averages of that type are common in the analysis of network or matching data. We then show how to apply our results to approximate joint distributions of means in several variables and when the statistic of interest is an estimator that is defined by potentially nonlinear moment conditions. Another extension is to non-exhaustively matched data, when not all of the $N\times T$ index pairs are observed. Finally we consider the case in which the $(i,t)$ index pairs correspond to clusters of more than one unit.

\subsection{Clustering in $D$ Dimensions}

The bootstrap procedure can be immediately extended to the case of an array $\left(Y_{i_1\dots i_D}:i_1=1,\dots,N_1,\dots,i_D=1,\dots,N_D\right)$ that may exhibit clustering in $D$ dimensions. As in the benchmark case, we assume that the sampling units corresponding to the indices in each dimension are i.i.d. draws from a common distribution so that for the $d$th dimension the ``planes" of the form $\left(Y_{i_1\dots i_{d-1}ji_{d+1}\dots i_D}:i_{d'}=1,\dots,N_{d'},d'\neq d\right)$ are i.i.d. for $j=1,\dots,N_d$.

From the main result by \cite{Hoo79}, a random array of this form can be represented as
\[Y_{i_1,\dots,i_D} = f(\alpha_{1i_1},\dots,\alpha_{Di_D},\varepsilon_{i_1\dots,i_D})\]
for some function $f(a_1,\dots,a_D,e)$, where $\alpha_{1i_1},\dots,\alpha_{Di_D},\varepsilon_{i_1\dots i_D}$ are i.i.d. draws for their respective marginal distributions, w.l.o.g. the uniform distribution.


For a sampling protocol of this type, we can adapt the bootstrap procedure from section \ref{sec:algorithm_sec} in a straightforward manner: Let
\[\bar{Y}_{N_1\dots N_D}:=\frac1{\prod_{d=1}^DN_d}\sum_{i_1,\dots,i_D}Y_{i_1\dots i_D}\]
and compute the projections of the array on the $d$th dimension,
\[\hat{a}_{di_d}:=\frac{N_d}{\prod_{d=1}^DN_d}\sum_{i_1,\dots,i_{d-1}\\ i_{d+1},\dots,i_D}Y_{i_1\dots i_D}-
\bar{Y}_{N_1\dots N_D}\]
and the projection residual
\[\hat{w}_{i_1\dots i_D}:=Y_{i_1\dots i_D} - \bar{Y}_{N_1\dots N_D} - \sum_{d=1}^D\hat{a}_{di_d}\]
For each dimension $d=1,\dots,D$, we draw $a_{di_d}^*$ independently from the empirical distribution for $\hat{a}_{di_d}$, and let $w_{i_1\dots i_D}^*:=\hat{w}_{k_1^*(i_1)\dots k_D^*(i_D)}\left(\prod_{d=1}^D\omega_{di_d}\right)$ for independent draws $\omega_{di_d}$ from the same distribution as in the baseline case. As before, $k_d^*(i_d)$ denotes the index of the cross-sectional unit corresponding to the $i_d$th bootstrap draw for dimension $d$. We then form
\[Y_{i_1\dots i_D}^*:=\bar{Y}_{N_1\dots N_D} + \sqrt{\hat{\lambda}}\sum_{d=1}^Da_{di_d}^* + w_{i_1\dots i_D}^*\]
with $\hat{\lambda}:=\frac{\max\left\{0,\sum_{d=1}^D\hat{\sigma}_{a_d}^2/N_d\right\}}
{\sum_{d=1}^D\hat{\sigma}_{a_d}^2/N_d
+D\hat{\sigma}_w^2/\prod_{d=1}^DN_d}$ defined in analogy to the two-dimensional case, and compute the bootstrapped mean $\bar{Y}_{N_1\dots N_D}^*:=\frac1{N_1\dots N_D}\sum_{i_1\dots i_D}Y_{i_1\dots i_D}^*$.

Noting that the arguments behind Theorems \ref{bootstrap_cons_thm} and \ref{refinement_thm} do not rely on the assumption that the random array is two-dimensional, an extension of these results to the $D$-dimensional case requires only a few minor notational changes.

\subsection{Dyadic and $D$-adic Data.}

The results in this paper readily extend to the case of dyadic or network data, where we observe a $D$-dimensional array $\left(Y_{i_1\dots i_D}:i_1,\dots,i_D=1,\dots,N\right)$ where a typical entry can be represented as
\[Y_{i_1\dots i_D} = f(\alpha_{i_1},\dots,\alpha_{i_D},\varepsilon_{i_1\dots i_D})\]
and $\alpha_1,\dots,\alpha_N$ and $\varepsilon_{i_1\dots i_D}$ are i.i.d. arrays. We can then consider the sampling distribution of the ``$D$-adic" mean
\[\bar{Y}_{N,D}:=\frac1{N^D}\sum_{i_1,\dots,i_D=1}^N Y_{i_1\dots i_D}\]
for $N$ units drawn at random from a larger population (with replacement) or distribution.\footnote{Note that the case in which we only include $D$-ads of $D$ or fewer distinct indices in the average is nested in this formulation, potentially after rescaling the mean by a bounded sequence.}

\begin{ex}\textbf{Subgraph Counts.} Suppose that the adjacency matrix with entries $G_{ij}\in\{0,1\}^{N^2}$ represents the subgraph among the set of nodes $1,\dots,N$ drawn at random from an infinite directed graph. Then the sampling distribution for the density of network homomorphisms (adjacency-preserving maps, see \cite{Lov12}) with respect to a network $F$ among $D$ distinct nodes can be approximated using this bootstrap procedure in the following way: We can define an indicator $R_{i_1\dots i_D}(F)$ that equals 1 if there is an adjacency-preserving map between $F$ and the subnetwork among the nodes $i_1,\dots,i_D$. We can then re-sample from the $D$-dimensional array with entries $Y_{i_1\dots,i_D}:=R_{i_1\dots i_D}(F)$ using the algorithm described above, where in step (b) we draw $N$ row identifiers with replacement at random and select columns and other dimensions of the array corresponding to the same identifiers.
\end{ex}

We can implement the bootstrap for $D$-adic arrays by following the algorithm as described in Section \ref{sec:algorithm_sec} except that in step (b) we draw $N$ row identifiers with replacement at random and select columns and other dimensions of the array corresponding to the same identifiers. The proofs of Theorems \ref{bootstrap_cons_thm} and \ref{refinement_thm} then go through under analogous conditions as for the original case.

\subsection{Multivariate Case}

Another important extension concerns the case of the mean of a vector-valued array $\bar{\mathbf{Y}}_{NT}=(\bar{Y}_{1NT},\dots,\bar{Y}_{MNT})'$, where $\mathbf{Y}_{it}\in\mathbb{R}^M$, and the joint distribution of the components of $\mathbf{Y}_{it}$ is left unrestricted. This generalization is relevant for joint tests and estimators that are defined by a vector of estimating equations described in the next subsection below.

For this case, we can consider  a component-wise Aldous-Hoover representation of the array
\[\mathbf{Y}_{it} = f(\mathbf{\alpha}_i,\mathbf{\gamma}_t,\mathbf{\varepsilon}_{it})\]
Here $\mathbf{\alpha}_i,\mathbf{\gamma}_t,\mathbf{\varepsilon}_{it}\in\mathbb{R}^M$ are i.i.d., but the individual components of the vectors $\mathbf{\alpha}_i$, $\mathbf{\gamma}_t$, and $\mathbf{\varepsilon}_{it}$, respectively, may be dependent in an arbitrary fashion.

We can then implement the bootstrap algorithm from the baseline case jointly in all $M$ components of the random vector $\mathbf{Y}_{it}$, where the projections $\mathbf{\hat{a}}_i$, $\mathbf{\hat{g}}_t$ and $\mathbf{\hat{w}}_{it}$ are $M$-dimensional vectors whose components are defined in analogy to the scalar case. The shrinkage parameters $\hat{\lambda}_1,\dots,\hat{\lambda}_M$ are then computed component by component as in the univariate case.

We denote the respective rates for the individual components with $\mathbf{r}_{NT}=(r_{1NT},\dots,r_{MNT})'$, where $r_{mNT}^2:=\var(\bar{Y}_{mNT})$, the variance of the $m$th component of the sample average $\bar{\mathbf{Y}}_{NT}$. We also denote slowest component of $\mathbf{r}_{NT}$ with $\varrho_{NT}:=\max_{m=1,\dots,M}|r_{mNT}|$. Then using the Cram\'er-Wold device, it follows immediately from Theorem \ref{bootstrap_cons_thm} that the bootstrap remains consistent for approximating the joint distribution of $\diag(\mathbf{r}_{NT})(\bar{\mathbf{Y}}_{NT}-\mathbb{E}[\mathbf{Y}_{it}])$ if the conditions of that theorem hold for each component $m=1,\dots,M$. Similarly, a refinement at the $\varrho_{NT}^{-2}$ rate is a straightforward extension of Theorem \ref{refinement_thm}.

\subsection{Bootstrapping Estimators}

The bootstrap procedure developed for the distribution of the sample mean $\bar{Y}_{NT}$ can be used to estimate the distribution of potentially nonlinear estimators. Specifically, suppose that the estimand of interest is a parameter $\theta_0$ in some parameter space $\Theta\subset\mathbb{R}^k$ which satisfies moment conditions of the form
\[\mathbb{E}[g(Y_{it};\theta_0)] = 0 \]
for a known function $g:\mathcal{Y}\times\Theta\rightarrow\mathbb{R}^m$. We can obtain a Z-estimator $\hat{\theta}$ for the parameter by solving $m$ estimating equations of the form
\[0 = \hat{A}_{NT}\hat{S}_{NT}(\hat{\theta})\]
where we define the score $\hat{S}_{NT}(\theta):=\frac1{NT}\sum_{i=1}^N\sum_{t=1}^T g(Y_{it};\theta)$,
and $\hat{A}_{NT}$ is an $k\times m$ matrix which may depend on quantities estimated from the data with probability limit $\hat{A}_{NT}\stackrel{p}{\rightarrow}A_0$. If we denote the Jacobian of the population moment with $G_0:=\nabla_{\theta}\mathbb{E}[g(Y_{it};\theta_0)]$, under regularity conditions we have from standard arguments\footnote{See e.g. \cite{NMc94}} that the estimator is asymptotically linear and satisfies the expansion
\[r_{NT}(\hat{\theta}-\theta_0) =  - \left(A_0G_0\right)^{-1}r_{NT}\hat{S}_{NT}(\theta_0)+o_p(1)\]
where $r_{NT}$ is a rate such that the distribution of $r_{NT}S_{NT}(\theta_0)$ is asymptotically tight.

Following the proposal by \cite{KlS12}, we can obtain the bootstrap analog $\hat{S}_{NT}^*(\hat{\theta}):=\frac1{NT}\sum_{i=1}^N g_{it}^*$ by resampling from the $N\times T\times m$ array with entries $g_{it}:=g(Y_{it};\hat{\theta})$ using the (multivariate version of the) algorithm from Section \ref{sec:algorithm_sec}. We can then estimate the distribution of the estimator with
\[r_{NT}(\hat{\theta}^*-\hat{\theta}):= - \left(\hat{A}_{NT}\hat{G}_{NT}\right)^{-1}r_{NT}\hat{S}_{NT}^*(\hat{\theta})\]
where $\hat{G}_{NT}:=\frac1{NT}\sum_{i=1}^N\sum_{t=1}^T \nabla_{\theta}g(Y_{it};\hat{\theta})$.
It is important to note that refinements are generally only available if the estimating equations are linear in the parameter, so that the estimator can be represented as a smooth function of sample moments.

An important special case are method of moments estimators that match model predictions as a function of the unknown parameter $\pi:\Theta\rightarrow\mathbb{R}^M$ to the corresponding sample moments, $\frac1{NT}\sum_{i=1}^N\sum_{t=1}^Tg(Y_{it})$. In that case, we can directly bootstrap the joint distribution of the corresponding scores of the form
\[ \hat{S}_{NT}(\theta) = \frac1{NT}\sum_{i=1}^N\sum_{t=1}^T(g(Y_{it}) - \pi(\theta))\]
Note that the resulting estimating equations are linear in the sample moments by construction, so that the bootstrap procedure immediately inherits the asymptotic properties from the bootstrap distribution for vectors of sample means, including refinements.

\subsection{Non-Exhaustively Matched Samples}

We next consider the case in which $Y_{it}$ is observed for a subset of index pairs $(i,t)$. For example, units $i=1,\dots,N$ could be high school students, and $t=1,\dots,T$ teachers, and we observe student $i$'s test score $Y_{it}$ after being taught by teacher $t$. For this problem, we could think of the sampling frame as follows: for each classroom, a group of students is selected according to some protocol, and at the same time a teacher is assigned to that classroom. The assignment process may be ``blind" to student and teacher-level characteristics $\alpha_i$ or $\gamma_t$, or subject to sorting. E.g. a principal may assign a more talented teacher to a classroom of ``weak" students. Sorting raises a number of conceptual and practical issues, and for the remainder of this section we restrict our attention to the case of ``random" assignment, as described by the following ``no sorting" condition below.

We can formalize such a sampling scheme by defining an $N\times T$ matrix $\mathbf{W}$ of indicator variables, where $W_{it}$ equals one if $Y_{it}$ is observed for the dyad $(i,t)$, and zero otherwise. We then consider the sampling distribution of
\[\bar{Y}_{NT,W}:=\frac1{\sum_{i=1}^N\sum_{t=1}^TW_{it}}\sum_{i=1}^N\sum_{t=1}^TW_{it}Y_{it}\]
conditional on $W_{it}$. We also define $T_i:=\sum_{t=1}^TW_{it}$ and $N_t:=\sum_{i=1}^NW_{it}$, and let
\[p_i:=\frac1T\sum_{t=1}^TW_{it},\;p_t:=\frac1N\sum_{i=1}^NW_{it},\;\textnormal{and }\bar{p}:=\frac1{NT}\sum_{i=1}^N\sum_{t=1}^TW_{it}=\frac1N\sum_{i=1}^Np_i=\frac1T\sum_{t=1}^T p_t\]
We then make the following assumptions:

\begin{ass}\label{nonexhaustive_ass} (a) As $N,T\rightarrow\infty$ sampling weights $W_{it}$ are such that $\frac1N\sum_{i=1}^N\left(p_i/\bar{p}\right)^2\rightarrow\kappa_a<\infty$ and $\frac1T\sum_{t=1}^T\left(p_t/\bar{p}\right)^2\rightarrow\kappa_g<\infty$. (b) The random array can be represented as $Y_{it}=h(\alpha_i,\gamma_t,\varepsilon_{it})$ for some function $h(\cdot)$, and random variables $\alpha_i,\gamma_t,\varepsilon_{it}$ that are i.i.d. conditional on $W_{it}$.
\end{ass}

Note that part (a) does not impose any restrictions on the density/sparseness of the sampling frame, but the assumption of finite limits $\kappa_a,\kappa_g$ amounts to a balance requirement on relative cluster sizes in either dimension. In particular we allow for the case $\bar{p}\rightarrow0$, but rule out the existence of individual clusters that dominate in size. Part(b) can be interpreted as a ``no sorting" condition that is restrictive in many contexts in which the observable dyads are the result of matching or self-selection of economic agents. This includes the cases of joint output measures for matched employer/employee data with assortative matching on worker and firm productivity, as well as test scores for samples of students and teachers if students and teachers are matched according to ability.

Given Assumption \ref{nonexhaustive_ass}, we find from elementary variance calculations that
\begin{eqnarray}\label{r_NTW_def}r_{NT,W}^2&:=&\var(\bar{Y}_{NT,W})\\
\nonumber&=&\frac1{NT\bar{p}}\left(T\bar{p}\sigma_a^2
\left[\frac1N\sum_{i=1}^N\left(\frac{p_i}{\bar{p}}\right)^2\right]
+ N\bar{p}\sigma_g^2\left[\frac1T\sum_{t=1}^T\left(\frac{p_t}{\bar{p}}\right)^2\right] + \sigma_w^2\right)\end{eqnarray}
From this expression, we can see that clustering on $\alpha_i$ and $\gamma_t$ matters asymptotically if and only if $N\bar{p}+T\bar{p}$ converges to a strictly positive limit. Cluster-level variation dominates the limiting distribution if $N\bar{p}+T\bar{p}\rightarrow\infty$.

By Assumption \ref{nonexhaustive_ass} (b), $\mathbb{E}[\bar{Y}_{NT,W}|\mathbf{W}]=\mathbb{E}[Y_{it}|W_{it}]=\mathbb{E}[Y_{it}]$ a.s., so that our analysis of the asymptotic distribution will focus on the studentized mean $r_{NT}(\bar{Y}_{NT,W} - \mathbb{E}[Y_{it}])$.

We then consider the following bootstrap algorithm:
\begin{itemize}
\item Generate an exhaustively matched bootstrap sample $Y_{it}^*$, $i=1,\dots,N$, $t=1,\dots,T$ as in the baseline case with \[\hat{\lambda}_{NT}:=\frac{(T\bar{p}-1)\hat{\sigma}_a^2
\left[\frac1N\sum_{i=1}^N\left(\frac{p_i}{\bar{p}}\right)^2\right]
+ (N\bar{p}-1)\hat{\sigma}_g^2\left[\frac1T\sum_{t=1}^T\left(\frac{p_t}{\bar{p}}\right)^2\right]}
{(T\bar{p}-1)\hat{\sigma}_a^2
\left[\frac1N\sum_{i=1}^N\left(\frac{p_i}{\bar{p}}\right)^2\right]
+ (N\bar{p}-1)\hat{\sigma}_g^2\left[\frac1T\sum_{t=1}^T\left(\frac{p_t}{\bar{p}}\right)^2\right] + 2\bar{p}\hat{\sigma}_w^2}.\]


\item Keep the observations for which $W_{it}=1$ and compute the bootstrapped mean
\[\bar{Y}_{NT,W}^*:=\frac1{\sum_{i=1}^N\sum_{t=1}^TW_{it}}\sum_{i=1}^N\sum_{t=1}^TW_{it}Y_{it}^*\]
\end{itemize}


We can then show that under Assumptions \ref{integ_ass} and \ref{nonexhaustive_ass}, the analogous conclusions to Theorems \ref{bootstrap_cons_thm} and \ref{refinement_thm} hold for the modified bootstrap distribution:

\begin{thm}\textbf{(Bootstrap Consistency)}\label{nonexhaust_cons_thm} Suppose that Assumptions \ref{integ_ass} and \ref{nonexhaustive_ass} hold. Then (a) the sampling distribution $\mathbb{P}_{NT}(r_{NT}(\bar{Y}_{NT,W}-\mathbb{E}[Y_{it}]))$ and the bootstrap distribution $\mathbb{P}_{NT}^*(r_{NT}(\bar{Y}_{NT,W}^*-\bar{Y}_{NT,W}))$ converge in probability to the same limit,
\[\|\mathbb{P}_{NT}^*(r_{NT}(\bar{Y}_{NT,W}-\mathbb{E}[Y_{it}]))-\mathbb{P}_{NT}(r_{NT}(\bar{Y}_{NT,W}^*
-\bar{Y}_{NT,W}))\|_{\infty}
\stackrel{p}{\rightarrow}0\]
Furthermore, (b) under Assumption \ref{spectral_ass} convergence is uniform.
\end{thm}

See the appendix for a proof. The only major complication arises if the second-order projection term $\frac1{NT\bar{p}^2}\sum_{i=1}^N\sum_{t=1}^TW_{it}[h(\alpha_i,\gamma_t)-a_i-g_t + \mathbb{E}[Y_{it}])$ remains relevant in the limit. In that case, the terms $\frac1{NT\bar{p}}\sum_{i=1}^N\sum_{t=1}^TW_{it}\phi_k(\alpha_i)\psi_k(\gamma_t)$ of the sparse representation can in general no longer be represented in terms of separate sample averages of $\phi_k(\alpha_i)$ and $\psi_k(\gamma_t)$, respectively. Instead we use results on random quadratic forms by \cite{GTi99} to reach the analogous conclusions.

Note that for clustering in more than two dimensions, our argument is only valid for the case $q_v=0$. For the case of a sparse sample, $\bar{p}\rightarrow0$, Corollary 2 in \cite{GTi99} furthermore implies the stronger conclusion of asymptotic normality of $r_{NT}(\bar{Y}_{NT}-\mathbb{E}[Y_{it}])$ even when $q_v>0$. Finally, a straightforward adaptation of the arguments in the proof of Theorem \ref{refinement_thm} establishes refinements to the estimated percentiles for the case of non-exhaustively matched samples whenever $q_v=0$.

\subsection{Unbalanced Cluster Sizes}


Suppose that we observe $R_{it}\in\mathbb{N}$ i.i.d. units in the intersection of clusters $i$ and $t$, denoted by $Y_{itr},\;r=1,\dots,R_{it}$. We consider inference for the average of pooled observations,
\[\bar{Y}_{NT,R}:=\frac1{\sum_{i=1}^N\sum_{t=1}^TR_{it}}\sum_{i=1}^N\sum_{t=1}^T\sum_{r=1}^{R_{it}} Y_{itr}\]
We also define $r_i:=\frac1T\sum_{t=1}^TR_{it}$, $r_t:=\frac1N\sum_{i=1}^NR_{it}$, and $\bar{r}:=\frac1{NT}\sum_{i=1}^N\sum_{t=1}^TR_{it}$. Clearly, $\bar{r}=\frac1N\sum_{i=1}^Nr_i = \frac1T\sum_{t=1}^T r_t$.

Note that for the case of equal-sized clusters, $R_{it}=R$, this problem is formally equivalent to clustering in three dimensions $i=1,\dots,N$, $t=1,\dots,T$, and $r=1,\dots,R$, where clustering in the third dimension is trivial, and the Aldous-Hoover representation is of the form
\[Y_{itr}=h(\alpha_i,\gamma_t,\varepsilon_{itr})\]
where $\alpha_i,\gamma_t,\varepsilon_{itr}$ are i.i.d. across all indices. Note that in the case of balanced cluster sizes, $R_{it}=R$ for all $i,t$, we can directly apply our results for the baseline case, where $Y_{it}:=\frac1R\sum_{r=1}^RY_{itr}$. The unbalanced case in which $R_{it}$ varies across $i,t$ requires additional assumptions under which we can adapt our approach for the case of non-exhaustively matched samples from the previous section. However, our results do not assume that $R$ grows large.

For our results we assume that cluster size is independent of cluster effects $\alpha_i,\gamma_t$, and that the imbalance in cluster size is bounded:
\begin{ass}\label{unbalanced_ass} (a) As $N,T\rightarrow\infty$ sampling weights $R_{it}$ are such that $\bar{r}\rightarrow\infty$, $\frac1N\sum_{i=1}^N\left(r_i/\bar{r}\right)^2\rightarrow\kappa_a<\infty$ and $\frac1T\sum_{t=1}^T\left(r_t/\bar{r}\right)^2\rightarrow\kappa_g<\infty$. (b) The random array satisfies $Y_{it}=h(\alpha_i,\gamma_t,\varepsilon_{it})$, where $\alpha_i,\gamma_t,\varepsilon_{it}$ are i.i.d. conditional on $R_{it}$.
\end{ass}

Let $\hat{a}_i:=\frac1{Tr_t}\sum_{t=1}^T\sum_{r=1}^{R_{it}}Y_{itr} - \bar{Y}_{NT,R}$ and $\hat{g}_t:=\frac1{Nr_t}\sum_{i=1}^N\sum_{r=1}^{R_{it}}Y_{itr} - \bar{Y}_{NT,R}$. Furthermore,
\[\hat{v}_{it}:=\frac1{R_{it}}\sum_{r=1}^{R_{it}}Y_{itr} - \hat{a}_i - \hat{g}_t+\bar{Y}_{NT,R}\]
and
\[\hat{e}_{itr}:=Y_{itr} - \hat{a}_i - \hat{g}_t - \hat{v}_{it}\]
For our projection representation, $\hat{v}_{it}$ estimates the second projection term $\mathbb{E}[Y_{itr}|\alpha_i,\gamma_t]$, and $\hat{e}_{itr}$ may remain relevant for the limiting distribution as long as $R$ does not grow too fast.

We then construct a bootstrap sample as follows:
\begin{itemize}
\item Generate $a_i^*:=\hat{a}_{k(i)}$, $g_t^*:=\hat{g}_{s(t)}$ for $k(i)$ and $s(t)$ drawn independently and uniformly at random from the index sets $\{1,\dots,N\}$ and $\{1,\dots,T\}$, respectively, and $v_{it}^*:=\hat{v}_{k(i)s(t)}$ and $e_{itr}^*:=\hat{e}_{k(i)s(t)q(r)}$ for $q(r)$ drawn independently and uniformly from $\{1,\dots, R_{k(i),s(t)}\}$.
\item Let $\omega_i,\omega_t,\omega_r$ be i.i.d draws from a distribution with mean zero, unit variance, and third moments equal to one for $i=1,\dots,N$, $t=1,\dots,T$, and $r=1,\dots,R$.
\item Generate an $N\times T\times R$ array of bootstrap draws
\[Y_{itr}^*:= \bar{Y}_{NT,R} + \sqrt{\hat{\lambda}}(a_i^* + g_t^*) + \omega_i\omega_t(\sqrt{\hat{\tau}}v_{it}^* +
\omega_re_{itr}^*)\]
where $\hat{\tau}:=\frac{(\bar{r}-1)\hat{\sigma}_v^2}{(\bar{r}-1)\hat{\sigma}_v^2 + \hat{\sigma}_e^2}$ and
\[\hat{\lambda}_{NT}:=\frac{(T\bar{r}-1)\hat{\sigma}_a^2
\left[\frac1N\sum_{i=1}^N\left(\frac{r_i}{\bar{r}}\right)^2\right]
+ (N\bar{r}-1)\hat{\sigma}_g^2\left[\frac1T\sum_{t=1}^T\left(\frac{r_t}{\bar{r}}\right)^2\right]}
{(T\bar{r}-1)\hat{\sigma}_a^2
\left[\frac1N\sum_{i=1}^N\left(\frac{r_i}{\bar{r}}\right)^2\right]
+ (N\bar{r}-1)\hat{\sigma}_g^2\left[\frac1T\sum_{t=1}^T\left(\frac{r_t}{\bar{r}}\right)^2\right] + 2\bar{r}\hat{\sigma}_w^2}\]
\end{itemize}

Under Assumptions \ref{integ_ass} and \ref{unbalanced_ass}, the analogous conclusions to Theorems \ref{bootstrap_cons_thm} and \ref{refinement_thm} regarding bootstrap consistency and refinements hold for the modified bootstrap procedure after only minor modifications of the arguments in Theorem \ref{nonexhaust_cons_thm}. Note that the restriction of the proof of Theorem \ref{nonexhaust_cons_thm} to the case of clustering in two dimensions only applies to the case $q_v>0$, for which refinements are not available even in the baseline setup.

\section{Simulation Study}
\label{sec:simulations}

We now present simulation results to demonstrate the performance of the bootstrap procedure. We consider balanced and unbalanced designs with additively separable and nonseparable cluster effects. Particular attention is given to the degenerate cases of uncorrelated observations, and drifting sequences. We report false rejection rates of the corresponding two-sided test of the null of a zero mean at a $5\%$ significance level for four alternative procedures
\begin{itemize}
\item[(GAU)] ``Plug-in" Gaussian inference using a consistent estimator $\hat{s}_{NT}^2$ for the asymptotic variance of $\bar{y}_{NT}$ that is robust to two-way clustering,
\item[(BS)] inference based on the bootstrap estimate for the distribution of $r_{NT}\bar{y}_{NT}$,
\item[(PIV)] inference based on the bootstrap estimate for the distribution of the studentized mean, $t_{NT}:=\left(\widehat{\var}(\bar{y}_{NT})\right)^{-1/2}\bar{y}_{NT}$,
\item[(SYM)] symmetric inference based on the bootstrap estimate for the distribution of the absolute value of the studentized mean, $|t_{NT}|$.
\end{itemize}
According to our theoretical results, any of these inference procedures is asymptotically valid, while PIV and SYM provide refinements over GAU and BS. It also follows from standard arguments (see e.g. \cite{Hor01}) that refinements from SYM should be of a higher order than those obtained for PIV. We also report the simulated bias for asymptotic variance estimation using (a) the analytic estimator in \cite{CGM11} AN and (b) the bootstrap variance estimator BS. Simulation results were obtained from 5000 simulated samples with bootstrap distributions approximated using $1000$ bootstrap draws. For all bootstrap results, we use a modification of the distribution for $\omega_i,\omega_t,\omega_{it}$ in the Wild bootstrap that corrects for finite-sample bias in the second and third moments in the empirical distribution. That modification is derived in Appendix \ref{sec:mom_corr_app}.

For the first set of results, we generate a two-way clustered array according to the additively separable design
\[y_{it} = \sigma_{\alpha}\alpha_i + \sigma_{\gamma}\gamma_t + \sigma_{\varepsilon}\varepsilon_{it}\]
where $\gamma_t,\varepsilon_{it}$ are i.i.d. standard normal. We generated $\alpha_i=(\zeta_i - \mu_{\alpha})/\kappa_{\alpha}$ for $\log\zeta_i\sim N(0,1)$, where $\mu_{\alpha}=\mathbb{E}[\zeta_i]$, and $\kappa_{\alpha}^2=\var(\alpha_i)$ were obtained using analytic formulae for the moments of the log-normal distribution. In particular, the distribution of $\alpha_i$ is skewed to the right.

Our simulation designs vary the relative importance of the three factors through the choice of  $\sigma_{\alpha},\sigma_{\gamma},\sigma_{\varepsilon}$. Design 1 (non-degenerate case) chooses $\sigma_{\alpha}^2=\sigma_{\gamma}^2=\sigma_{\varepsilon}^2=1$, Design 2 (degenerate case) sets $\sigma_{\alpha}^2=\sigma_{\gamma}^2=0$ and $\sigma_{\varepsilon}^2=1$. Design 3 considers the drifting sequence $\sigma_{\alpha}^2=5/T$, $\sigma_{\gamma}^2 = 5/N$, and $\sigma_{\varepsilon}^2=1$.

In order to illustrate the quality of the approximation for the separable cases, we report rejection rates for the bootstrap procedure based on the estimator $\hat{\lambda}$ for the variance ratio, which is uniformly valid except when $q_v>0$. False rejection rates based on the conservative version combining the procedures with estimators $\hat{\lambda}$ and $\tilde{\lambda}$ can by design not exceed those reported here, but may be significantly lower in some cases.

Results for the balanced case are given in Table \ref{table:mc_bal_des} and largely support our theoretical claims. In particular, for all four procedures rejection rates approach the nominal $0.05$ significance level as $N$ and $T$ grow. For Design 1, PIV and SYM show a marked improvement over GAU and BS which is consistent with asymptotic refinements established in Theorem \ref{refinement_thm}. These improvements are more pronounced for one-sided than two-sided rejection rates. We can see from the simulation results that the respective biases in estimating percentiles in the lower and upper tails of the distribution via GAU or BS have opposite signs, so that these biases partially offset each other for two-sided tests. For Design 2, our theoretical results imply not refinements for PIV and SYM since for that specification, $y_{it}=\sigma_{\varepsilon}\varepsilon_{it}$ is i.i.d. Gaussian. Design 3 considers drifting sequences of DGPs for which Theorem \ref{refinement_thm} does not predict refinements.

\begin{table}\scriptsize{\begin{tabular}{rcccccccccccccccccc}
&&&\multicolumn{2}{c}{Estimated Variance}&&\multicolumn{4}{c}{FRR, Two-Sided}&&
\multicolumn{3}{c}{FRR, One-Sided (L)}&&\multicolumn{3}{c}{FRR, One-Sided (R)}\\\cline{4-5}\cline{7-10}\cline{12-14}\cline{16-18}\\[-4pt]
$N$&$T$&\hspace{0.3cm}&AN&BS&\hspace{0.1cm}&GAU&BS&PIV&SYM&\hspace{0.1cm}&GAU&BS&PIV
&&GAU&BS&PIV\\[2pt]
\cline{4-5}\cline{7-10}\cline{12-14}\cline{16-18}\\
&&&\multicolumn{14}{c}{Design 1}\\[2pt]\cline{4-18}\\
10&10 && 1.042 & 1.164 && 0.074 & 0.064 & 0.057 & 0.049&& 0.091 & 0.080 & 0.069 &&0.033&0.027 &0.043\\
20&20 && 1.048 & 1.099 && 0.066 & 0.058 & 0.062 & 0.051&& 0.080 & 0.077 & 0.067 &&0.030&0.026 &0.045\\
50&50 && 1.032 & 1.052 && 0.053 & 0.049 & 0.059 & 0.048&& 0.071 & 0.071 & 0.062 &&0.035&0.030 &0.049\\
100&100 && 0.982 & 0.993 && 0.057 & 0.052 & 0.061 & 0.052&& 0.070 & 0.072 & 0.061 &&0.037&0.033 &0.052\\[4pt]
&&&\multicolumn{14}{c}{Design 2}\\[2pt]\cline{4-18}\\
10&10 && 1.306 & 1.365 && 0.041 & 0.039 & 0.047 & 0.046&& 0.040 & 0.039 & 0.043 &&0.040&0.039 &0.046\\
20&20 && 1.225 & 1.243 && 0.038 & 0.035 & 0.042 & 0.041&& 0.039 & 0.038 & 0.042 &&0.040&0.039 &0.046\\
50&50 && 1.142 & 1.146 && 0.042 & 0.042 & 0.044 & 0.044&& 0.041 & 0.040 & 0.043 &&0.042&0.043 &0.045\\
100&100 && 1.051 & 1.052 && 0.048 & 0.048 & 0.051 & 0.052&& 0.053 & 0.051 & 0.055 &&0.048&0.049 &0.051\\[4pt]
&&&\multicolumn{14}{c}{Design 3}\\[2pt]\cline{4-18}\\
10&10 && 1.166 & 1.237 && 0.055 & 0.054 & 0.061 & 0.057&& 0.050 & 0.048 & 0.053 &&0.050&0.047 &0.056\\
20&20 && 1.099 & 1.124 && 0.055 & 0.053 & 0.059 & 0.056&& 0.051 & 0.050 & 0.053 &&0.044&0.044 &0.051\\
50&50 && 1.008 & 1.014 && 0.058 & 0.057 & 0.061 & 0.059&& 0.060 & 0.059 & 0.062 &&0.049&0.049 &0.051\\
100&100 && 1.014 & 1.017 && 0.052 & 0.052 & 0.052 & 0.052&& 0.054 & 0.054 & 0.055 &&0.048&0.048 &0.050\\
&\vspace{-0.2cm}\\\cline{4-18}\\
\end{tabular}
\caption{Balanced separable case: Mean ratio of analytical (AN) and bootstrap (BS) estimators of asymptotic variance over sampling variance, false rejection rates (FRR) for two-sided and one-sided tests of the null $\mathbb{E}[Y_{it}]=0$. Design 1: $\sigma_a^2=\sigma_g^2 =\sigma_e^2=1$; Design 2: $\sigma_a^2=\sigma_g^2=0,\sigma_e^2=1$; Design 3: $\sigma_a^2=1/T,\sigma_g^2=1/N,\sigma_e^2=1$.}
\label{table:mc_bal_des}}
\end{table}

\begin{figure}\footnotesize\label{fig:balanced_rep}
\includegraphics[scale=0.5]{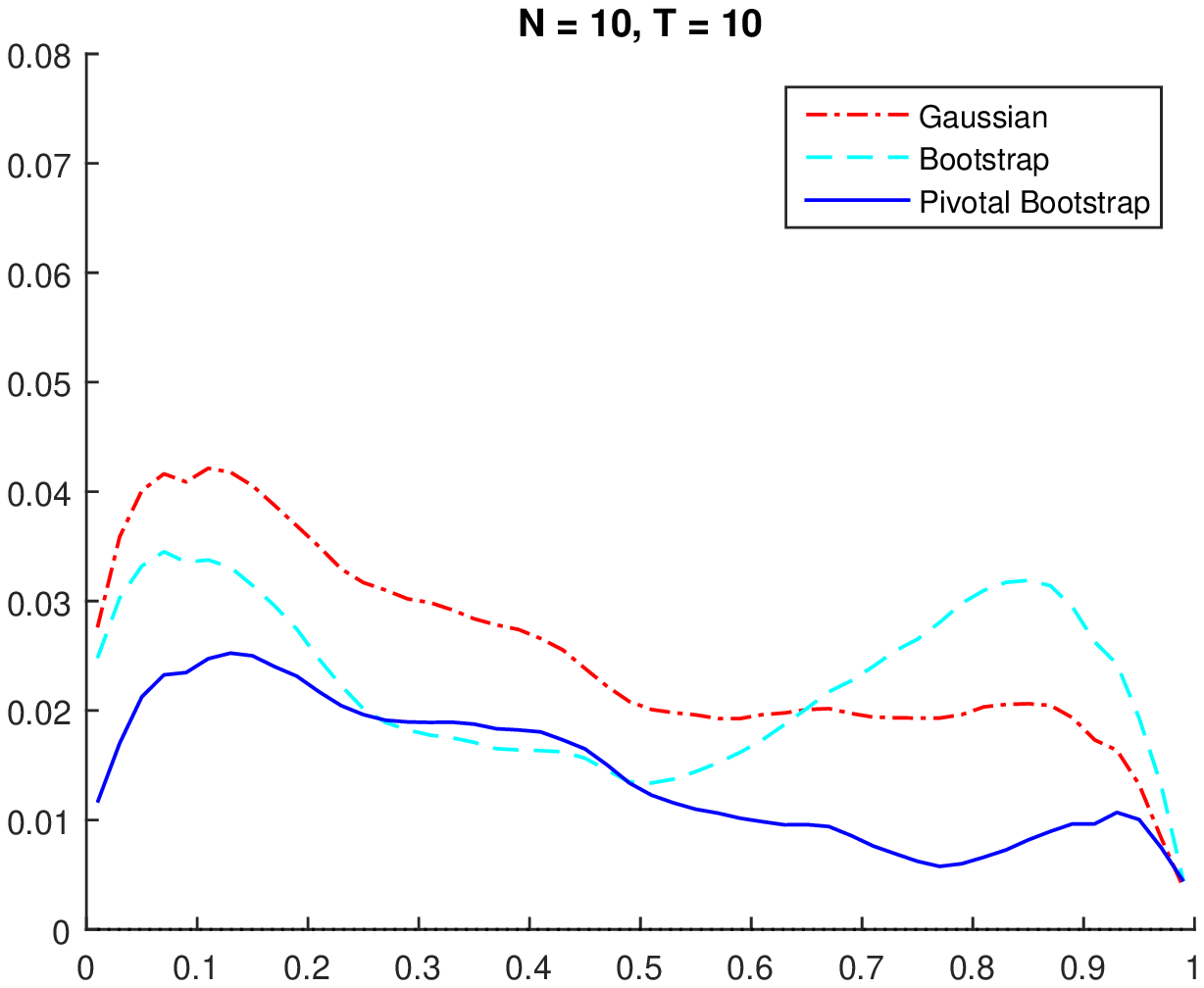}\includegraphics[scale=0.5]{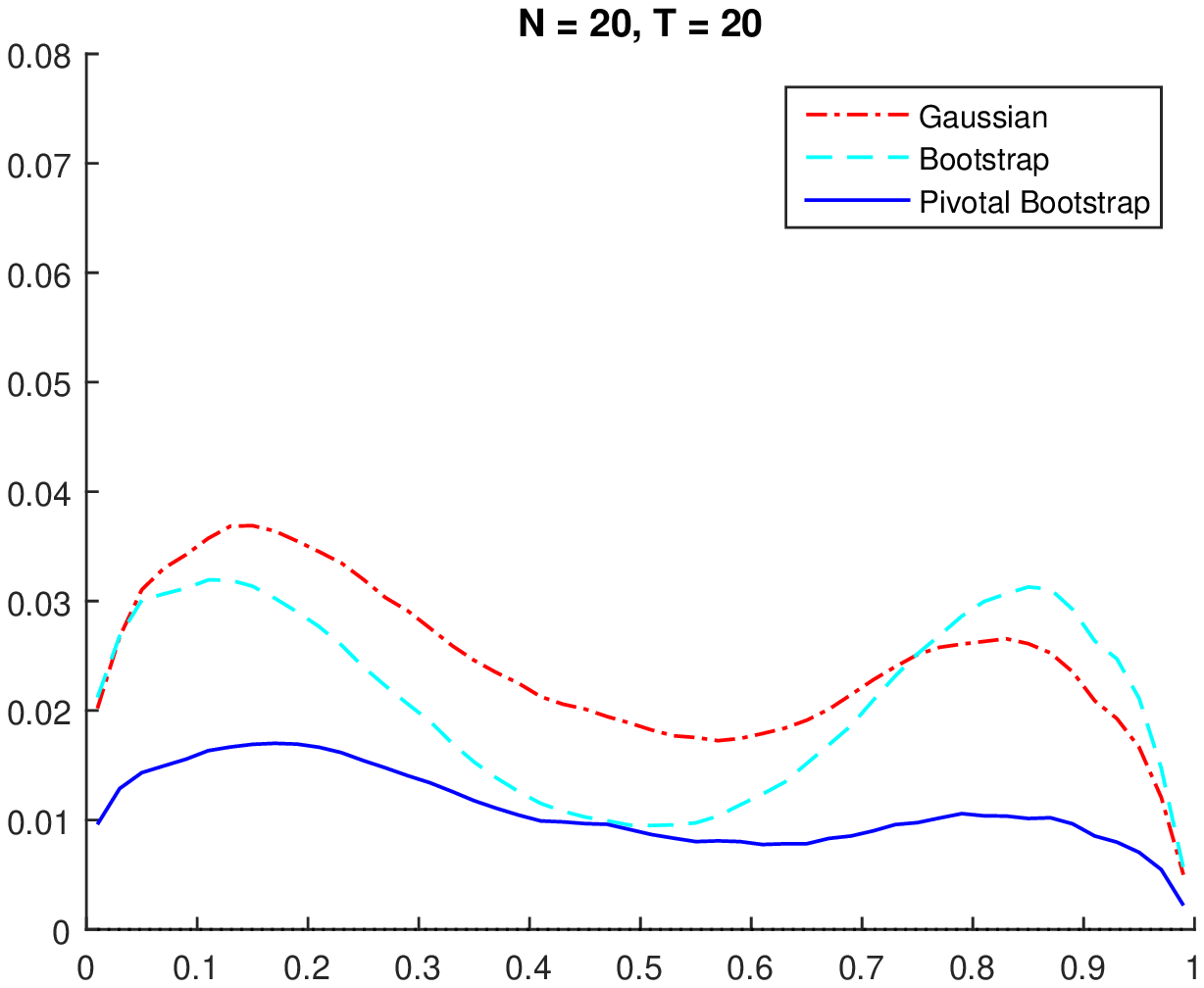}\\
\includegraphics[scale=0.5]{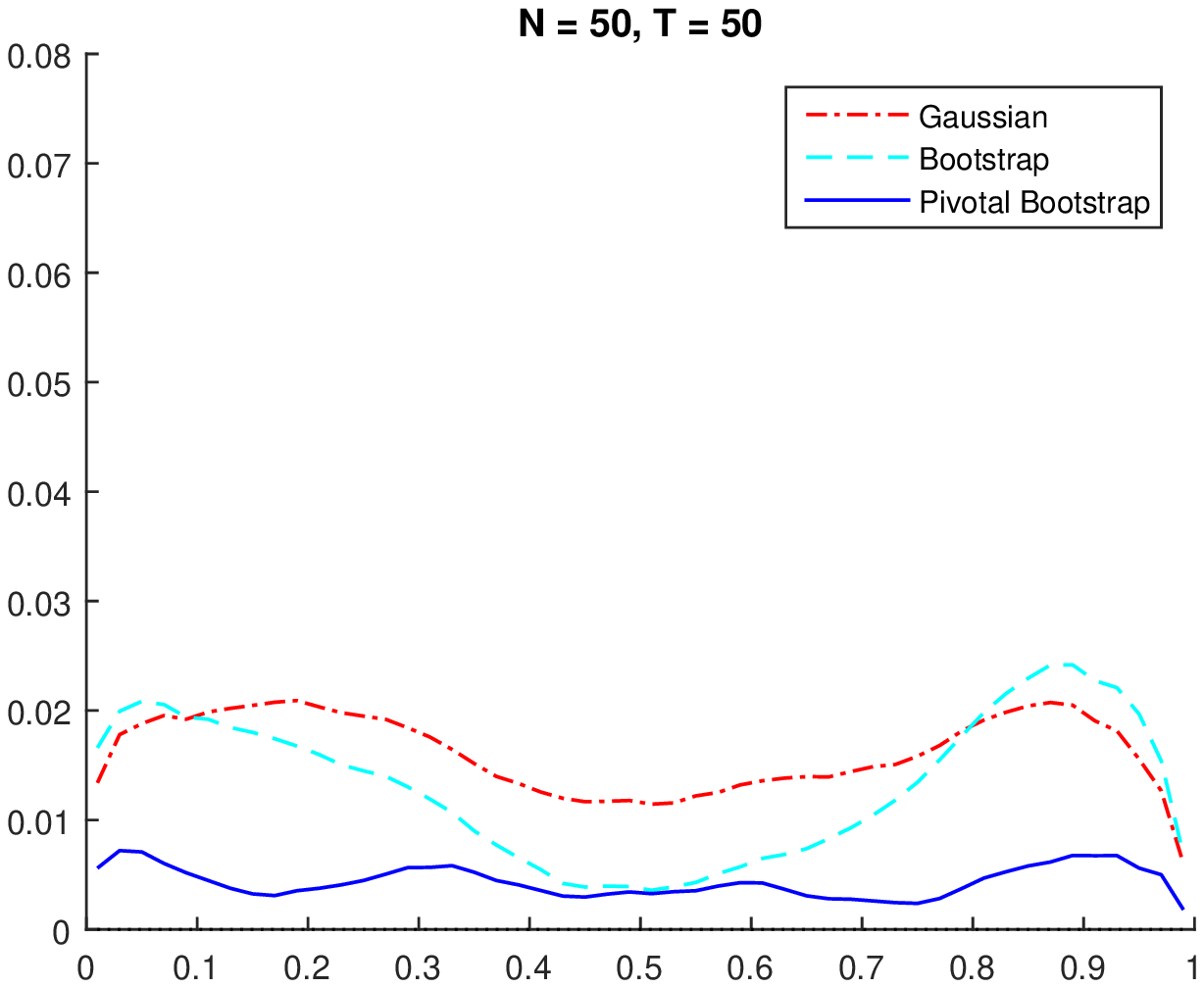}\includegraphics[scale=0.5]{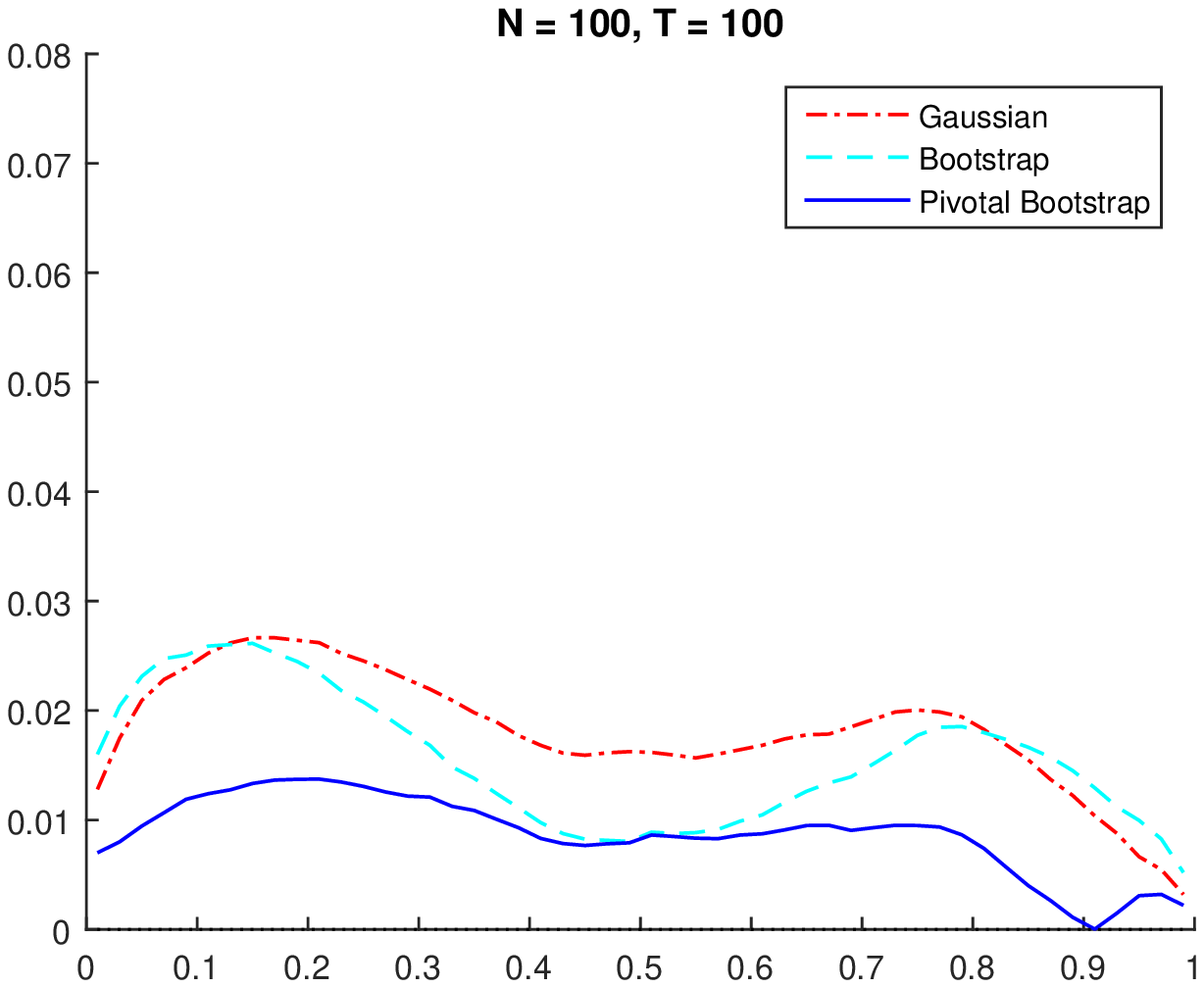}
\caption{Balanced separable case: Absolute error in estimated c.d.f., plotted against nominal percentiles. Plots are based on Design 1: $\sigma_a^2=1,\sigma_g^2 =0.2,\sigma_e^2=1$.}
\end{figure}

We also simulate the absolute error in rejection probabilities based on GAU, BS, and PIV at all percentiles for Design 1. Specifically, we estimate the percentiles of the sampling distribution for each simulated sample using either method, and simulate the frequency at which the t-statistic for the sample exceeds each percentile. Figure \ref{fig:balanced_rep} reports the absolute difference between the simulated and nominal rejection frequencies. We find that for all three methods, the absolute discrepancy between nominal and simulated rejection rates decreases as $N$ and $T$ grow across all percentiles. The standard bootstrap (BS) does not exhibit a clear improvement relative to plug-in asymptotic approximation (GAU), whereas rejection rates based on the bootstrap for the studentized mean (PIV) are consistently closer to nominal levels. We report additional results for percentiles relevant for one- and two-sided tests at commonly used significance levels in the appendix.

We next assess the importance of balance in the relative sizes of $N$ and $T$, as well as the relative importance of clustering in either dimension. In particular, we first consider balanced designs $T=N$ where we set $\sigma_a=1,\sigma_g=0.5$ and $\sigma_e=1$. We then consider unbalanced designs where we let $N=10,20,50,100$ vary while holding $T=20$ fixed, see Table \ref{table:mc_unbal_des} for simulation results. While the bootstrap is not asymptotically valid if $T$ remains fixed, results are broadly in line with those for the balanced case for the cooresponding sample size. Overall, these results are again consistent with theoretical predictions on asymptotic validity and refinements.

\begin{table}\scriptsize{\begin{tabular}{rcccccccccccccccccccc}
&&&\multicolumn{2}{c}{Estimated Variance}&&\multicolumn{4}{c}{FRR, Two-Sided}&&
\multicolumn{3}{c}{FRR, One-Sided (L)}&&\multicolumn{3}{c}{FRR, One-Sided (R)}\\\cline{4-5}\cline{7-10}\cline{12-14}\cline{16-18}\\[-4pt]
$N$&$T$&\hspace{0.3cm}&AN&BS&\hspace{0.1cm}&GAU&BS&PIV&SYM&\hspace{0.1cm}&GAU&BS&PIV
&&GAU&BS&PIV\\[2pt]
\cline{4-5}\cline{7-10}\cline{12-14}\cline{16-18}\\
&&&\multicolumn{14}{c}{Design 1}\\[2pt]\cline{4-18}\\
10&10 && 1.033 & 1.149 && 0.082 & 0.071 & 0.063 & 0.057&& 0.097 & 0.089 & 0.073 &&0.035&0.029 &0.039\\
20&20 && 1.047 & 1.100 && 0.067 & 0.063 & 0.059 & 0.051&& 0.087 & 0.084 & 0.066 &&0.031&0.026 &0.041\\
50&50 && 1.037 & 1.058 && 0.056 & 0.057 & 0.051 & 0.051&& 0.071 & 0.073 & 0.059 &&0.032&0.026 &0.046\\
100&100 && 0.965 & 0.976 && 0.061 & 0.059 & 0.062 & 0.056&& 0.072 & 0.073 & 0.058 &&0.042&0.039 &0.053\\[4pt]
&&&\multicolumn{14}{c}{Design 2}\\[2pt]\cline{4-18}\\
10&20 && 1.062 & 1.161 && 0.067 & 0.058 & 0.052 & 0.049&& 0.080 & 0.077 & 0.066 &&0.033&0.026 &0.038\\
20&20 && 0.971 & 1.023 && 0.067 & 0.059 & 0.061 & 0.055&& 0.071 & 0.068 & 0.059 &&0.046&0.040 &0.053\\
50&20 && 0.979 & 1.021 && 0.061 & 0.057 & 0.057 & 0.052&& 0.063 & 0.060 & 0.056 &&0.052&0.048 &0.051\\
100&20 && 1.004 & 1.050 && 0.063 & 0.055 & 0.051 & 0.048&& 0.060 & 0.058 & 0.053 &&0.055&0.052 &0.049\\[4pt]
&\vspace{-0.2cm}\\\cline{4-18}\\
\end{tabular}
\caption{Unbalanced separable case: Mean ratio of analytical (AN) and bootstrap (BS) estimators of asymptotic variance over sampling variance, false rejection rates (FRR) for two-sided and one-sided tests of the null $\mathbb{E}[Y_{it}]=0$. Design 1: $\sigma_a^2=0.5,\sigma_g^2 =0.1,\sigma_e^2=0.5$; Design 2: $\sigma_a^2=0.5,\sigma_g^2=0.5,\sigma_e^2=1$.}
\label{table:mc_unbal_des}}
\end{table}

Finally, we simulate a model with non-separable cluster effects, where we specify
\[y_{it} = (\alpha_i + \mu_{\alpha})(\gamma_t + \mu_{\gamma}) -\mu_{\alpha}\mu_{\gamma} + \varepsilon_{it}\]
for i.i.d. standard normal random variables $\alpha_i,\gamma_t$ and $\varepsilon_{it}$. We consider one non-degenerate design with $\mu_{\alpha}=\mu_{\gamma}=1$, and an alternative design with $\mu_{\alpha}=\mu_{\gamma}=0$ for which $y_{it}$ is not clustered in means, see Table \ref{table:mc_nonsep_des} for simulation results.

In preliminary simulation results we find that estimation error in $\hat{\lambda}$, shifting the relative weight of the normally distributed and the Wiener chaos component, affects the bootstrap estimates even for intermediate sample sizes $N,T=20,50$. Since we are primarily interested in illustrating the performance of the approximations, we therefore report results based on a pointwise consistent bootstrap procedure which uses the estimated variance ratio $\tilde{\lambda}$ with shrinkage towards zero. Specifically, we consider an adaptation of the procedure which replaces the variance estimates in $\hat{\lambda}_{NT}$ with $\hat{\sigma}_{a}^2\dum\{\hat{\sigma}_{a}^2>0.5\log(N)/\sqrt{N}\}$ and $\hat{\sigma}_{g}^2\dum\{\hat{\sigma}_{g}^2>0.5\log(T)/\sqrt{T}\}$ can be shown to be valid point-wise, although not uniformly, which is supported by the last set of simulation results in Table \ref{table:mc_nonsep_des}.

\begin{table}\scriptsize{\begin{tabular}{rcccccccccccccccccc}
&&&\multicolumn{2}{c}{Estimated Variance}&&\multicolumn{4}{c}{FRR, Two-Sided}&&
\multicolumn{3}{c}{FRR, One-Sided (L)}&&\multicolumn{3}{c}{FRR, One-Sided (R)}\\\cline{4-5}\cline{7-10}\cline{12-14}\cline{16-18}\\[-4pt]
$N$&$T$&\hspace{0.3cm}&AN&BS&\hspace{0.1cm}&GAU&BS&PIV&SYM&\hspace{0.1cm}&GAU&BS&PIV
&&GAU&BS&PIV\\[2pt]
\cline{4-5}\cline{7-10}\cline{12-14}\cline{16-18}\\
&&&\multicolumn{14}{c}{Design 1}\\[2pt]\cline{4-18}\\
10&10 && 1.018 & 0.926 && 0.091 & 0.104 & 0.052 & 0.052&& 0.111 & 0.119 & 0.078 &&0.031&0.039 &0.012\\
20&20 && 1.003 & 0.956 && 0.083 & 0.089 & 0.058 & 0.058&& 0.091 & 0.094 & 0.077 &&0.032&0.037 &0.023\\
50&50 && 0.997 & 0.978 && 0.061 & 0.064 & 0.053 & 0.054&& 0.075 & 0.076 & 0.068 &&0.038&0.040 &0.035\\
100&100 && 1.000 & 0.990 && 0.054 & 0.056 & 0.050 & 0.051&& 0.067 & 0.068 & 0.064 &&0.041&0.043 &0.040\\[4pt]
&&&\multicolumn{14}{c}{Design 2}\\[2pt]\cline{4-18}\\
10&10 && 1.304 & 1.248 && 0.046 & 0.039 & 0.080 & 0.079&& 0.040 & 0.044 & 0.067 &&0.040&0.043 &0.064\\
20&20 && 1.243 & 1.225 && 0.042 & 0.037 & 0.055 & 0.053&& 0.032 & 0.037 & 0.050 &&0.037&0.043 &0.054\\
50&50 && 1.146 & 1.141 && 0.041 & 0.038 & 0.048 & 0.047&& 0.040 & 0.044 & 0.051 &&0.036&0.041 &0.046\\
100&100 && 1.104 & 1.101 && 0.045 & 0.041 & 0.045 & 0.043&& 0.041 & 0.044 & 0.046 &&0.041&0.047 &0.050\\[4pt]
&\vspace{-0.2cm}\\\cline{4-18}\\
\end{tabular}
\caption{Non-separable case: Mean ratio of analytical (AN) and bootstrap (BS) estimators of asymptotic variance over sampling variance, false rejection rates (FRR) for two-sided and one-sided tests of the null $\mathbb{E}[Y_{it}]=0$. Design 1: $\sigma_a^2=0.5,\sigma_g^2 =0.5,\sigma_e^2=0.5$ and $\mu_a=\mu_g=1$; Design 2: $\sigma_a^2=0.5,\sigma_g^2=0.5,\sigma_e^2=0.1$ and $\mu_a=\mu_g=0$.}
\label{table:mc_nonsep_des}}
\end{table}

We find that in the non-degenerate case $\mu_{\alpha},\mu_{\gamma}\neq0$ the bootstrap produces results that are comparable to the separable case. According to our theoretical results, all four procedures are asymptotically valid, whereas PIV and SYM should produce refinements, which is consistent with the first set of simulation results. For the degenerate case, $\mu_{\alpha}=\mu_{\gamma}=0$, theory predicts that Gaussian inference is not asymptotically valid even when a consistent estimator of the asymptotic variance is used.

As for the separable case, we also simulate the absolute error in rejection probabilities based on GAU, BS, and PIV at all percentiles for the degenerate case in Design 2. For the plug-in asymptotic approximation based on the Gaussian distribution there is no clear sign of convergence, and based on the theoretical properties, bias in rejection rates should be expected to persist for arbitrarily large sample sizes. Also, since the studentized mean is not asymptotically pivotal in this scenario, theory also does not predict refinements for PIV or SYM. This is reflected in the simulation results, showing no systematic difference between the two bootstrap estimates, BS and PIV.

\begin{figure}\footnotesize
\includegraphics[scale=0.5]{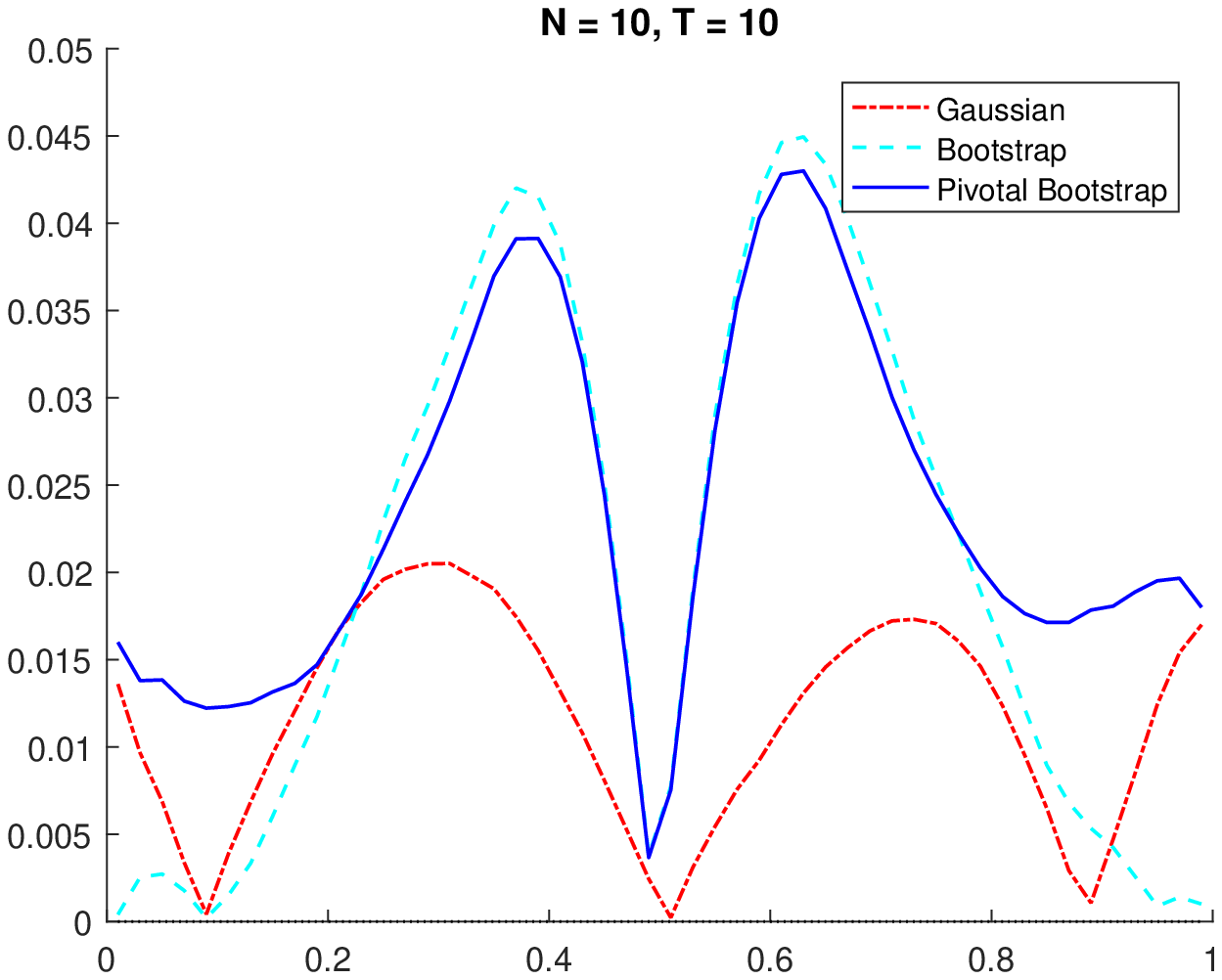}\includegraphics[scale=0.5]{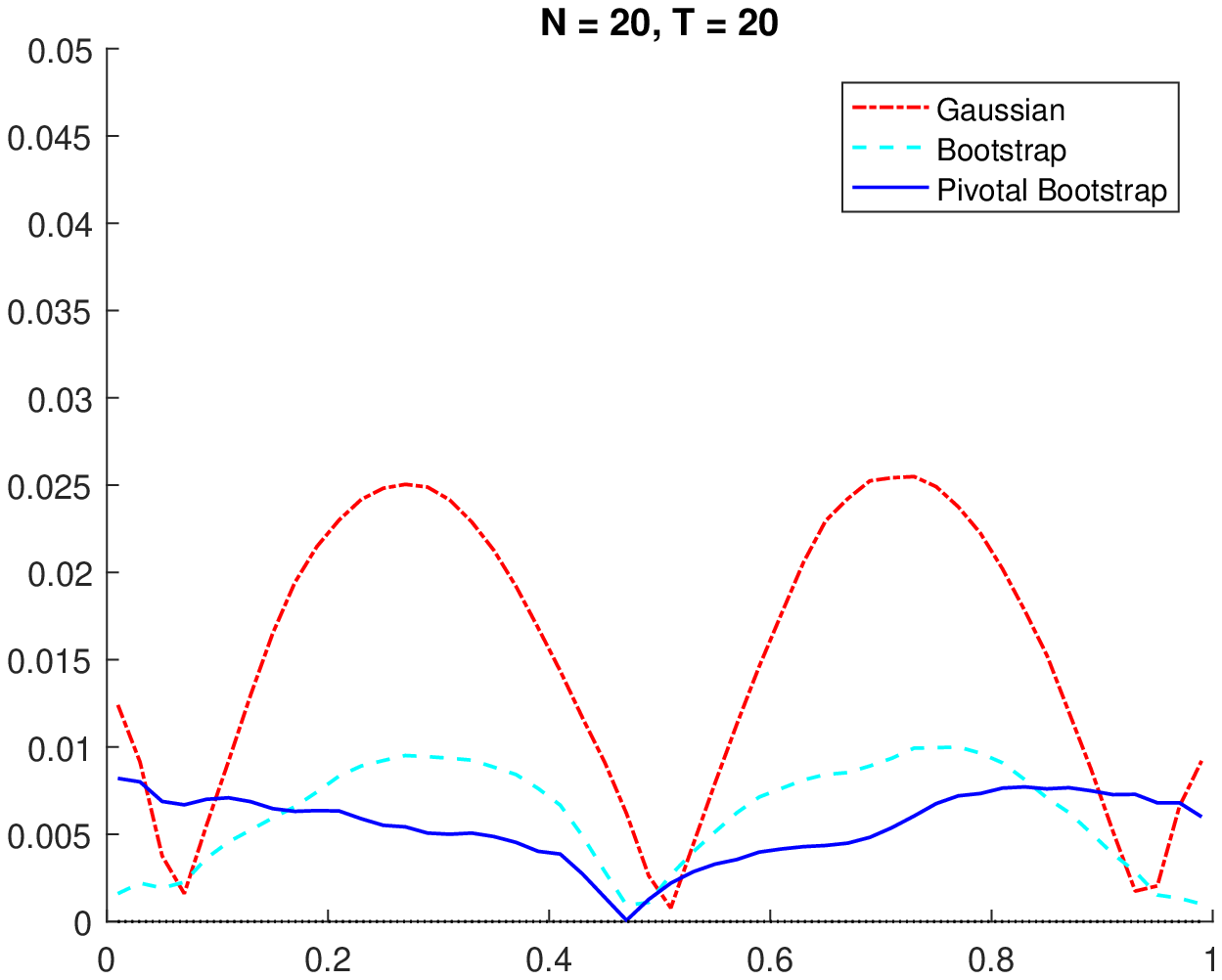}\\
\includegraphics[scale=0.5]{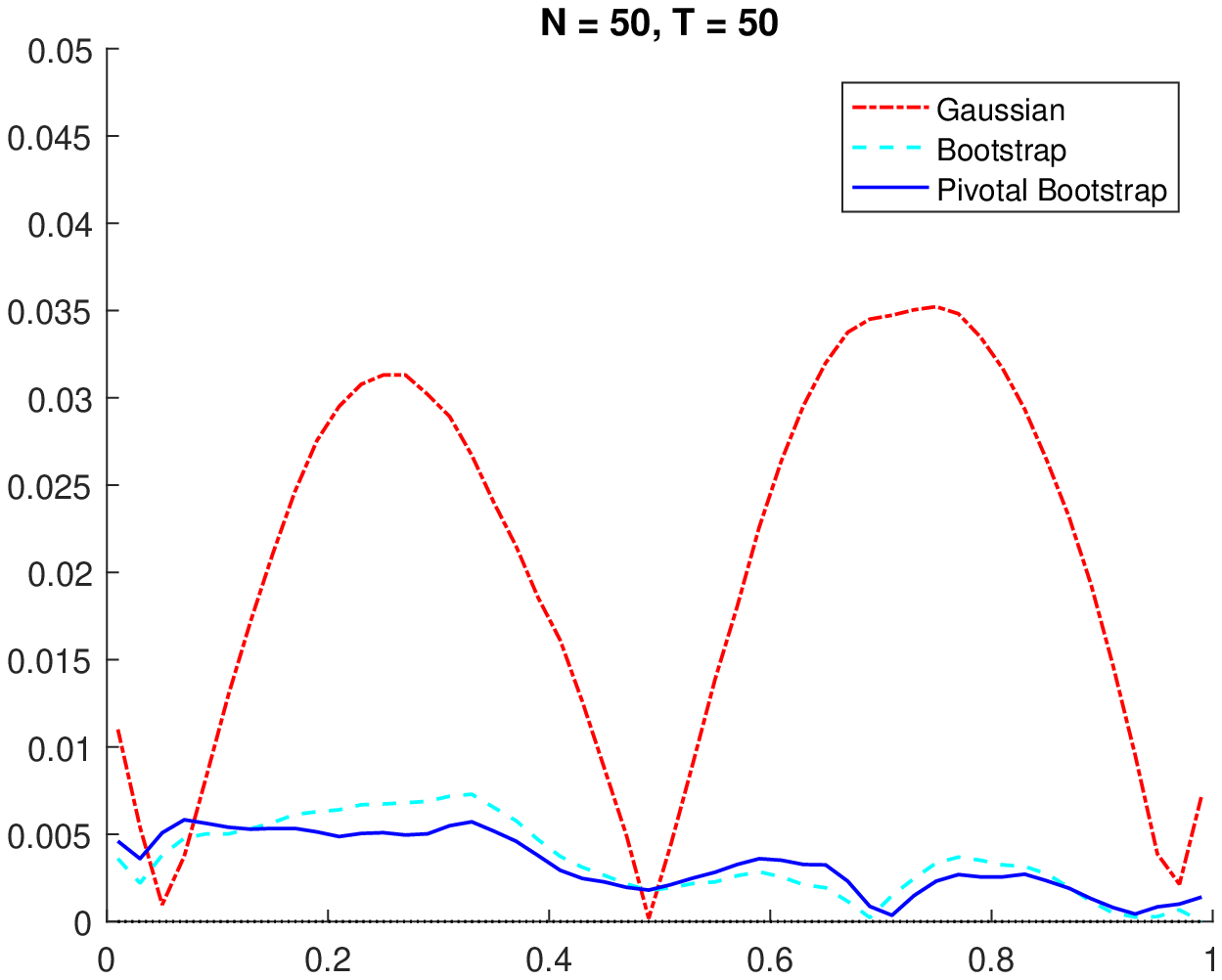}\includegraphics[scale=0.5]{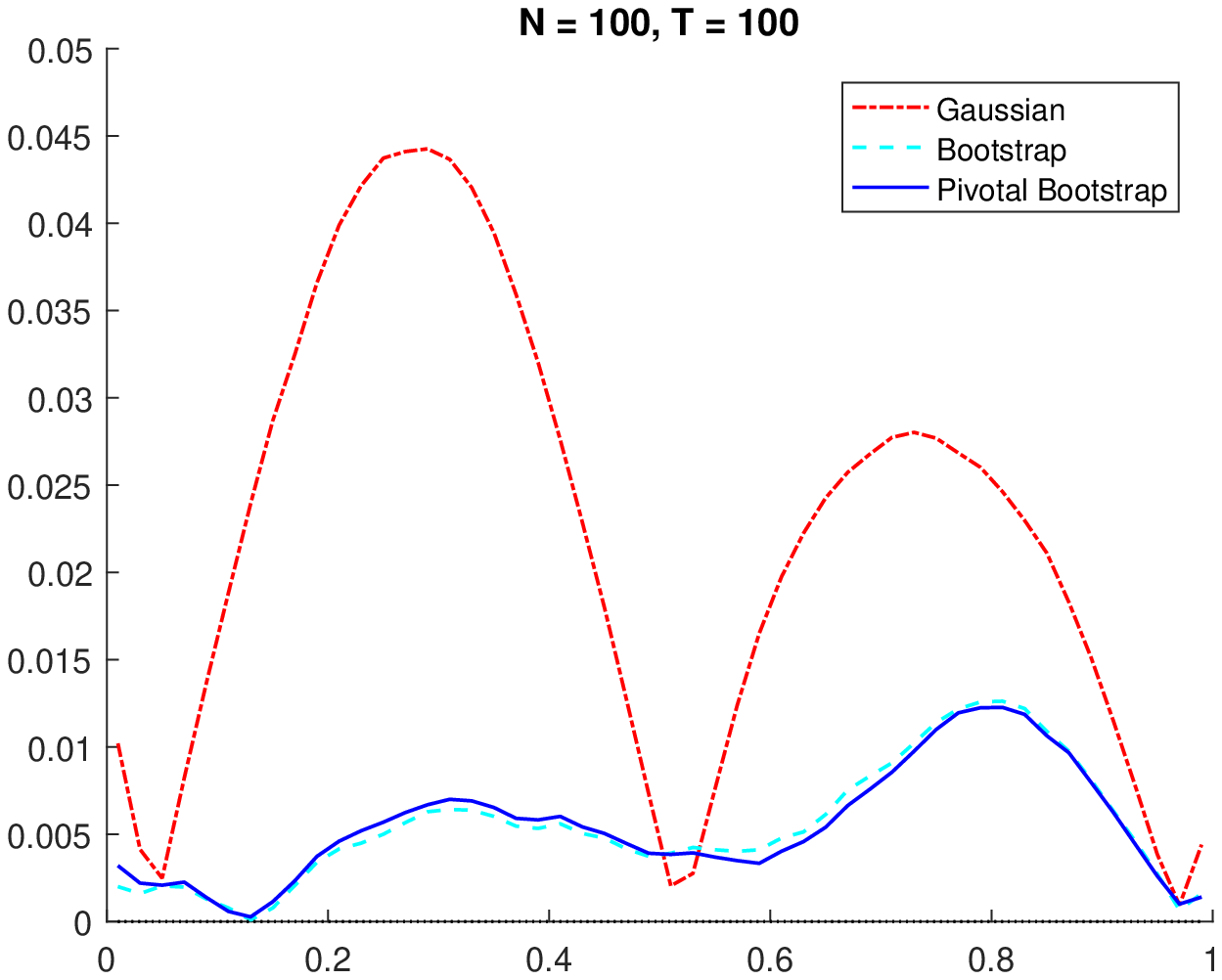}
\caption{Nonseparable case: Absolute error in estimated c.d.f., plotted against nominal percentiles. Plots are based on Design 2: $\sigma_a^2=0.5,\sigma_g^2=0.5,\sigma_e^2=0.1$ and $\mu_a=\mu_g=0$.}
\end{figure}


\footnotesize
\appendix

\section{Correcting Moments of Empirical Distribution}
\label{sec:mom_corr_app}

The second and third moments of the empirical distribution are both biased towards zero as estimators for the corresponding population moments. This bias vanishes asymptotically, but we find that the performance of the bootstrap improves for small and intermediate values of $N$ and $T$ if we replace the empirical distribution with a bias-corrected estimator. We first describe the general principle for that correction in terms of i.i.d. data and then show how to incorporate it into the procedure with multi-way clustering.

From a few straightforward calculations, we can see that for a sample of $N$ i.i.d. random variables $z_1,\dots,z_N$, the second and third central moments of the empirical distribution relate to their population analogs via
\[\mathbb{E}\left[\frac1N\sum_{i=1}^N(z_i-\bar{z}_N)^2\right] = \frac{N-1}N\mathbb{E}[(z_i-\mathbb{E}[z_i])^2]\;\textnormal{ and }\mathbb{E}\left[\frac1N\sum_{i=1}^N(z_i-\bar{z}_N)^3\right] =\frac{(N-2)(N-1)}{N^2}\mathbb{E}[(z_i-\mathbb{E}[z_i])^3]\]
where $\bar{z}_N:=\frac1N\sum_{i=1}^Nz_i$ is the sample mean.

We can now use the Wild bootstrap to obtain a distribution whose first three moments match those of the population in expectations. Specifically, we consider the conditional distribution of $z_i^*:=\omega_iz_i$ given the sample $z_1,\dots,z_N$, where the random variable $\omega_i$ is binary
\[\omega_i =\left\{\begin{array}{lcl}w_1&\hspace{0.3cm}&\textnormal{with probability }p\\
w_2&&\textnormal{with probability }1-p\end{array}\right.\]
and i.i.d. conditional on $z_1,\dots,z_n$. Adapting the proposal by \cite{Mam92}, we then choose the constants $w_1,w_2$ and $p$ subject to the moment conditions
\[\mathbb{E}[\omega_i] =0,\hspace{0.3cm}\mathbb{E}[\omega_i^2]=c_2,\hspace{0.3cm}\textnormal{and }\mathbb{E}[\omega_i^3]=c_3\]
for constants $c_2,c_3$ to be determined later. Up to a permutation, this system of moment conditions is solved by
\[p^*=\frac12-\frac12\sqrt{\frac{c_3^2}{4c_2^3 + c_3^2}},\hspace{0.2cm} w_1=\sqrt{\frac{1-p^*}{p^*}c_2},\hspace{0.3cm}\textnormal{and }
w_2=-\sqrt{\frac{p^*}{1-p^*}c_2}\]
Note that for $c_2=c_3=1$, we obtain the two-point distribution proposed for the Wild bootstrap by \cite{Mam92}, whereas for the Wild bootstrap correcting for bias in the first three moments in the empirical distribution, we choose $c_2 = \frac{N}{N-1}$ and $c_3 =\frac{N^2}{(N-2)(N-1)}$.

In order to implement our bootstrap procedure for multi-way clustering, we can resample each projection component $a_i,g_t,e_{it}$ separately using the adjusted Wild bootstrap. To that end, we choose $c_2,c_3$ as the analogous sequences in $N,T$ for the projection on either dimension, and $NT$ for second- and higher-order projection terms.

\section{Proofs}

\subsection*{Proof of Theorem \ref{sample_clt_thm}}

We can rewrite the projection terms in terms of the low-rank representation in \ref{spectral_rep}
\begin{eqnarray}
\nonumber e_{it}&=&Y_{it} - h(\alpha_i,\gamma_t)\\
\nonumber a_i&=&\mathbb{E}[h(\alpha_i,\gamma_t)|\alpha_i] - \mathbb{E}[h(\alpha_i,\gamma_t)] = \sum_{k=1}^{\infty}c_k\mathbb{E}[\psi_k(\gamma_t)](\phi_k(\alpha_i)-\mathbb{E}[\phi_k(\alpha_i)])\\
\nonumber g_t&=&\mathbb{E}[h(\alpha_i,\gamma_t)|\gamma_t] - \mathbb{E}[h(\alpha_i,\gamma_t)] = \sum_{k=1}^{\infty}c_k\mathbb{E}[\phi_k(\alpha_i)](\psi_k(\gamma_t)-\mathbb{E}[\psi_k(\gamma_t)])\\
\nonumber v_{it}&=&h(\alpha_i,\gamma_t) - a_i - g_t + \mathbb{E}[h(\alpha_i,\gamma_t)]\\
\nonumber &=& \sum_{k=1}^{\infty}c_k(\psi_k(\gamma_t) - \mathbb{E}[\psi_k(\gamma_t)])(\phi_k(\alpha_i)-\mathbb{E}[\phi_k(\gamma_t)])
\end{eqnarray}

%
%

Now let
\[\hat{Z}_N^a:=\frac1{N}\sum_{i=1}^Na_i,\hspace{0.3cm}\hat{Z}_T^g:=\frac1T\sum_{t=1}^Tg_t,\hspace{0.3cm}\textnormal{and }\hat{Z}_{NT}^e:=\frac1{NT}\sum_{i=1}^N\sum_{t=1}^T e_{it}\] By independence of $\alpha_i$ and $\gamma_t$, $\hat{Z}_N^a$ and $\hat{Z}_T^g$ are uncorrelated. We also define
\[\hat{Z}_{Nk}^{\phi}:=\frac1{\sqrt{N}}\sum_{i=1}^N(\phi_k(\alpha_i)-\mathbb{E}[\phi_k(\alpha_i)]),\hspace{0.5cm}
\hat{Z}_{Tk}^{\psi}:=\frac1{\sqrt{T}}\sum_{t=1}^T(\psi_k(\gamma_t)-\mathbb{E}[\psi_k(\gamma_t)])\]
for $k=1,2,\dots$. Since $\alpha_i$ and $\gamma_t$ are independent, $\hat{Z}_{Nk}^{\phi}$ and
$\hat{Z}_{Tk'}^{\psi}$ are uncorrelated for any pair $k,k'$. Also by orthogonality of the basis functions,
$\hat{Z}_{Nk}^{\phi}$ and $\hat{Z}_{Nk'}^{\phi}$ ($\hat{Z}_{Tk}^{\psi}$ and $\hat{Z}_{Tk'}^{\psi}$, respectively) are uncorrelated for any $k\neq k'$.  Finally by the projection properties of $a_i,g_t,e_{it}$ and $h(\alpha_i,\gamma_t)$, all remaining pairwise covariances among $\hat{Z}_N^a,\hat{Z}_T^g,\hat{Z}_{NT}^e$ and $\sum_{k=1}^{\infty}\hat{Z}_{Nk}^{\phi}\hat{Z}_{Tk}^{\psi}$ are zero.

We can stack these sample moments  \[\hat{Z}_{NT,K}:=\left(\hat{Z}_N^a,\hat{Z}_T^g,\hat{Z}_{NT}^e,\hat{Z}_{N1}^{\phi},\hat{Z}{T1}^{\psi},\dots,
\hat{Z}_{NK}^{\phi},\hat{Z}_{TK}^{\psi}\right)\]
so that by a multivariate CLT,
\[r_{NT}^{-1}\hat{Z}_{NT,K}\rightsquigarrow N(0,Q)\]
where $Q$ is a $(2K+3)\times(2K+3)$ matrix whose first three diagonal entries are $q_a,q_g,q_e$, and the remaining $2K$ diagonal entries are equal to $1$.

Truncating the expansion (\ref{spectral_rep}) at $K<\infty$, we define
\begin{eqnarray}\nonumber \bar{Y}_{NT,K}
\nonumber&=&b+\hat{Z}_N^a + \hat{Z}_N^g + \hat{Z}_{NT}^e + \sum_{k=1}^Kc_k\hat{Z}_{Nk}^{\phi}\hat{Z}_{Tk}^{\psi}
\end{eqnarray}
From the previous steps it then follows that
\[ r_{NT}^{-1}(\bar{Y}_{NT,K}-\mathbb{E}[Y_{it,K}])\rightsquigarrow
\sqrt{q_a + q_g + q_e}Z_0 + \sqrt{q_v}V_K\]
along each converging sequence, where $V_K:=\lim_{N,T}\frac1{\sigma_v}\sum_{k=1}^Kc_kZ_k^{\psi}Z_k^{\phi}$ with the coefficients $c_k$ potentially variying along the limiting sequence, and $Z_0,Z_1^{\phi},Z_1^{\psi},\dots,Z_K^{\phi},Z_K^{\psi}$ are i.i.d. standard normal random variables.

Note that convergence is uniform for every $K<\infty$, which establishes finite-dimensional convergence. Finally, notice that the approximation error with respect to the distribution of $r_{NT}^{-1}(\bar{Y}_{NT}-\mathbb{E}[Y_{it}])$ from the truncation at $K<\infty$ can be made arbitrarily small by choosing $K$ sufficiently large, where the magnitude of the approximation error can be controlled uniformly under Assumption \ref{spectral_ass}\qed


%

\subsection*{Proof of Theorem \ref{bootstrap_clt_thm}}

We can decompose $\hat{w}_{it}=\hat{v}_{it} + \hat{e}_{it}$, where
\begin{eqnarray}
\nonumber \hat{e}_{it}&=& e_{it}-\mathbb{E}_{NT}[e_{it}|\alpha_i] - \mathbb{E}_{NT}[e_{it}|\gamma_t] + \mathbb{E}_{NT}[e_{it}]\\
\nonumber \hat{v}_{it}&=& h(\alpha_i,\gamma_t) - \mathbb{E}_{NT}[h(\alpha_i,\gamma_t)|\alpha_i] - \mathbb{E}_{NT}[h(\alpha_i,\gamma_t)|\gamma_t] + \mathbb{E}_{NT}[h(\alpha_i,\gamma_t)]\\
\nonumber&=&\sum_{k=1}^{\infty}c_k(\psi_k(\gamma_t) - \mathbb{E}_{NT}[\psi_k(\gamma_t)])(\phi_k(\alpha_i)-\mathbb{E}_{NT}[\phi_k(\gamma_t)])
\end{eqnarray}

We then define the bootstrap analog of the process $\hat{Z}_{NT,K}$: Let
\[\hat{Z}_N^{a,*}:=\frac1{N}\sum_{i=1}^N\hat{a}_{j(i)},\hspace{0.3cm}
\hat{Z}_T^{g,*}:=\frac1T\sum_{t=1}^T\hat{g}_{s(t)},\hspace{0.3cm}\textnormal{and }\hat{Z}_{NT}^{e,*}:=\frac1{NT}\sum_{i=1}^N\sum_{t=1}^T\omega_{1i}\omega_{2t} \hat{e}_{j(i)s(t)}\]
Furthermore,
\[\hat{Z}_{Nk}^{\phi,*}:=\frac1{\sqrt{N}}\sum_{i=1}^N\omega_{1i}(\phi_k(\alpha_{j(i)})
-\mathbb{E}_{NT}[\phi_k(\alpha_i)]),
\hspace{0.5cm}
\hat{Z}_{Tk}^{\psi,*}:=\frac1{\sqrt{T}}\sum_{t=1}^T\omega_{2t}(\psi_k(\gamma_{s(t)})-\mathbb{E}_{NT}[\psi_k(\gamma_t)]\]
for $k=1,2,\dots$.

We can now combine these definitions for the bootstrap process to obtain
\begin{eqnarray}
\nonumber \bar{Y}_{NT}^*&:=&\bar{Y}_{NT} + \sqrt{\hat{\lambda}}\left(\frac1T\hat{Z}_N^{a,*} + \frac1N\hat{Z}_T^{g,*}\right)
+ \left(\hat{Z}_{NT}^{e,*} +\sum_{k=1}^{\infty}c_k\hat{Z}_{Nk}^{\phi,*}\hat{Z}_{Tk}^{\psi,*}\right)
\end{eqnarray}

Next notice that the first two moments of the bootstrap processes \[\left\{\frac{}{}\phi_k(\alpha_i)-\mathbb{E}_{NT}[\phi_k(\alpha_i)],
\psi_k(\gamma_t)-\mathbb{E}_{NT}[\psi_k(\gamma_t)]\right\}_{k=1}^K\]
under the empirical distribution converge in probability to their population analogs following standard arguments. It is also straightforward to verify that $\var(\hat{a}_i) = \sigma_a^2 + \frac1T\sigma_w^2$ and $\var(\hat{g}_t)=\sigma_g^2 + \frac1N\sigma_w^2$. In particular, $r_{NT}^{-2}\var(\hat{a}_i)\rightarrow q_a + q_w$, and  $r_{NT}^{-2}\var(\hat{g}_t)\rightarrow q_g + q_w$.


By Assumption \ref{integ_ass}, the third moments of $\hat{a}_i,\hat{\gamma}_t,\hat{e}_{it}$ under the empirical distribution are almost surely bounded, so that from the same argument as in the proof of Theorem 1 in \cite{Liu88}, the Berry-Es\'een theorem together with the Cram\'er-Wold device implies that
\[\hat{Z}_{NT}^*\rightsquigarrow N(0,Q^*)\]
conditional on $\left(Y_{it}\right)_{i=1,\dots,N\\t=1,\dots,T}$ almost surely. The first two diagonal elements of the asymptotic variance matrix $Q^*$ are given by $Q_{11}^* = q_a + q_w$, $Q_{22}^* = q_g + q_w$, and all other entries coincide with those of $Q$.

Now by assumption, the bootstrap procedure uses an estimator for $\lambda=\frac{q_a+q_g}{q_a+q_g+2q_w}$ that is consistent along the relevant parameter sequence. For any fixed $K<\infty$ we can therefore evaluate the limit of the truncated version of the bootstrapped statistic,
\begin{eqnarray}
\nonumber r_{NT}^{-1}(\bar{Y}_{NT,K}^*-\mathbb{E}_{NT}[Y_{it,K}])&\rightsquigarrow& 
\sqrt{q_a + q_g + q_e}Z^*
+\sqrt{q_v}V_K^*
\end{eqnarray}
where $V_K^*:=\lim_{N,T}\frac1{\sigma_v}\sum_{k=1}^Kc_kZ_k^{\psi,*}Z_k^{\phi,*}$, and $Z^*$ is a standard normal variable, independent of $V_K^*$. By standard approximation arguments, the distribution of $r_{NT}^{-1}(\bar{Y}_{NT,K}^*-\mathbb{E}_{NT}[Y_{it,K}])$ can then be approximated arbitrarily closely by choosing $K$ large enough, where the approximation error can be controlled uniformly under Assumption \ref{spectral_ass}\qed

\subsection*{Proof of Lemma \ref{var_comp_rate_lem}}

For part (a), let $\hat{s}_a^2:=\frac1{N-1}\sum_{i=1}^N\hat{a}_i^2$, $\hat{s}_g^2:=\frac1{T-1}\sum_{t=1}^T\hat{g}_t^2$, and $\hat{s}_w^2:=\frac1{NT-N-T}\sum_{i=1}^M\sum_{t=1}^T\hat{w}_{it}^2$ be the empirical variances of the projection terms $\hat{a}_i,\hat{g}_t,\hat{w}_{it}$. We can also verify that $\frac{N}{N-1}\var(\hat{a}_i)=\sigma_a^2 + \sigma_w^2/T$, $\frac{T}{T-1}\var(\hat{g}_t)=\sigma_g^2 + \sigma_w^2/N$, and $\frac{NT}{NT-N-T}\var(\hat{w}_{it}) = \sigma_w^2$.

Consider first the term $\hat{s}_a^2$: We can write
\[\hat{a}_i^2 = \left(a_i + \frac1T\sum_{t=1}^Tw_{it}\right)^2=
\left(a_i + \frac1T\sum_{t=1}^Te_{it}\right)^2 + 2\left(a_i + \frac1T\sum_{t=1}^Te_{it}\right)\frac1T\sum_{t=1}^Tv_{it} +  \left(\frac1T\sum_{t=1}^Tv_{it}\right)^2\]
Hence we have that
\begin{eqnarray}
\nonumber \hat{s}_a^2 - \left(\sigma_a^2+\frac1T\sigma_w^2\right)&=&
\frac1N\sum_{i=1}^N\left\{\left(a_i + \frac1T\sum_{t=1}^Te_{it}\right)^2-\left(\sigma_a^2+\frac1T\sigma_e^2\right)\right\}\\
\nonumber&&+\frac1{N}\sum_{i=1}^N\left(a_i + \frac1T\sum_{t=1}^Te_{it}\right)\frac1T\sum_{t=1}^Tv_{it}+\frac1{N}\sum_{i=1}^N
\left\{\left(\frac1T\sum_{t=1}^Tv_{it}\right)^2-\frac1T\sigma_v^2\right\}\\
\nonumber&=:&T_1 + T_2 + T_3
\end{eqnarray}
By independence and Lyapunov's CLT, we have that
\[T_1=O_P\left(N^{-1/2}\left(\sigma_a+T^{-1/2}\sigma_e\right)^2\right)\]
as $N\rightarrow\infty$. Next, consider the term $T_3$: defining $\tilde{\phi}_{ik}:=\phi_k(\alpha_i)-
\mathbb{E}[\phi_k(\alpha_i)]$ we can write


\begin{eqnarray}\nonumber\frac1{N}\sum_{i=1}^n\left(\frac1T\sum_{t=1}^Tv_{it}\right)^2&=&\frac1N\sum_{i=1}^N
\left(\frac1T\sum_{t=1}^T\sum_{k=1}^{\infty}c_k\tilde{\phi}_{ik}\tilde{\psi}_{tk}\right)^2\\
\nonumber &=&\frac1N\sum_{i=1}^N\sum_{k,k'}c_kc_{k'}\tilde{\phi}_{ik}\tilde{\phi}_{ik'}
\left(\sum_{t=1}^T\tilde{\psi}_{tk}\right)\left(\sum_{t=1}^T\tilde{\psi}_{tk'}\right)\\
\nonumber&=&\sum_{k,k'}c_kc_{k'}\left(\frac1N\sum_{i=1}^N\tilde{\phi}_{ik}\tilde{\phi}_{ik'}\right)
\left(\sum_{t=1}^T\tilde{\psi}_{tk}\right)\left(\sum_{t=1}^T\tilde{\psi}_{tk'}\right)\\
\label{emp_var_order_vit_app}&=:&\frac1T\sum_{k,k'}\left(\dum\{k=k'\} + \frac1{\sqrt{N}} \hat{Z}_{Nkk'}^{\tilde{\phi}\tilde{\phi}}\right)\hat{Z}_{Tk}^{\tilde{\psi}}\hat{Z}_{Tk'}^{\tilde{\psi}}
\end{eqnarray}
Here, $\hat{Z}_{Nkk'}^{\phi\phi}=\frac1{\sqrt{N}}\sum_{i=1}^N(\tilde{\phi}_{ik}\tilde{\phi}_{ik'}-\mathbb{E}[\tilde{\phi}_{ik}
\tilde{\phi}_{ik'}])$, where $\mathbb{E}[\tilde{\phi}_{ik}\tilde{\phi}_{ik'}]$ equals 1 if $k=k'$ and zero otherwise. In particular, it follows that
\[T_3 = O_P\left(T^{-1}\sigma_v^2\right)\]
as $N$ and $T$ grow large. By similar calculations, we find that
\begin{eqnarray}
\nonumber T_2&=&\sum_{k=1}^{\infty}c_k\left(\frac1N\sum_{i=1}^N\left(a_i + \frac1T\sum_{t=1}^Te_{it}\right)\tilde{\phi}_{ik}\right)\left(\frac1T\sum_{t=1}^T\tilde{\psi}_{tk}\right)\\
\nonumber&=&O_P\left(N^{-1/2}(\sigma_a + T^{-1/2}\sigma_e)T^{-1/2}\sigma_v\right)
\end{eqnarray}
noting that by construction $\mathbb{E}\left[a_i\tilde{\phi}_{ik}\right]=0$ for each $k=1,2,\dots$. Aggregating the contributions of the individual terms $T_1,T_2,t_3$, we then obtain
\[\hat{s}_a^2-\left(\sigma_a^2+\frac1T\sigma_w^2\right) = O_P\left(N^{-1/2}\left(\sigma_a+T^{-1/2}\sigma_e\right)^2 + T^{-1}\sigma_v^2\right)\]
Similarly, we find that
\[\hat{s}_g^2-\left(\sigma_g^2+\frac1N\sigma_w^2\right)=O_P\left(T^{-1/2}\left(\sigma_g+ N^{-1/2}\sigma_e\right)+N^{-1}\sigma_v^2\right)\]
Next, note that
\[\hat{\sigma}_w^2 = \frac1{NT}\sum_{i=1}^N\sum_{t=1}^T(v_{it}^2 + 2v_{it}e_{it} + e_{it}^2)\]
From calculations analogous to (\ref{emp_var_order_vit_app}), we also find that
\[\frac1{NT}\sum_{i=1}^N\sum_{t=1}^T v_{it}^2 = O_p\left(N^{-1/2}+T^{-1/2}\right)\]
Hence,
\[\hat{\sigma}_w^2 - \sigma_w^2 = O_P\left((NT)^{-1/2}\sigma_e^2 + (T^{-1/2} + N^{-1/2})\sigma_v^2\right)\]
The rates asserted in the Lemma then follow directly from the definitions of the variance estimators $\hat{\sigma}_a^2:=\hat{s}_a^2 - \frac1T\hat{s}_w^2$, $\hat{\sigma}_g^2:=\hat{s}_g^2 - \frac1N\hat{\sigma}_w^2$.

For a proof of part (b), note first that it is sufficient to find a specific family of distributions under which that rate cannot be improved upon. Specifically, consider the model
\[y_{it} = \alpha_i\gamma_t + \varepsilon_{it}\]
where $\alpha_i,\gamma_t,\varepsilon_{it}$ are independent, $\alpha_i\sim N(\mu_a,1)$, $\gamma_t\sim N(\mu_g,1)$ for some $\mu_a,\mu_g\geq0$, and $\varepsilon_{it}\sim N(0,\sigma_{\varepsilon}^2)$.

To establish the rate for the contribution of terms depending on $\sigma_v^2$ to that bound, consider the case $\sigma_{\varepsilon}^2=0$ and $\mu_a=0$. For this model, $a_i:=\mathbb{E}[y_{it}|\alpha_i]=\alpha_i\mu_g$ and $v_{it} = \alpha_i(\gamma_t-\mu_g)$, so that $\sigma_a^2 = \mu_g^2$ and $\sigma_v^2 = 1$. Clearly, $\mu_g$ cannot be estimated from the original data at a better rate than from directly observing $(\alpha_i)_{i=1}^N$ and $(\gamma_t)_{t=1}^T$. Furthermore, since $\gamma_1,\dots,\gamma_T$ are i.i.d., there exists no consistent test for the problem $H_0:\mu_g=0$ against $H_1:\mu_g=T^{-1/2}m$ for some $m>0$. Since under $H_0$, $\sigma_a^2=0$, whereas under $H_1$, $\sigma_a^2=T^{-1}m$, there can be no estimator for $\sigma_a^2$ that is consistent at a rate faster than $T^{-1}\sigma_v^2$.

The respective contributions of terms depending on $\sigma_a^2,\sigma_g^2$ and $\sigma_e^2$ to the rate bound follow immediately from standard arguments for the case of i.i.d. data, which can similarly be cast in terms of pairwise testing problems between drifting DGP sequences. Finally, consistent estimation of $\sigma_a^2$ under all DGPs permitted by our framework requires simultaneously solving these pairwise testing problems that gave us the respective rate contributions depending on $\sigma_a^2,\sigma_g^2,\sigma_e^2$ and $\sigma_v^2$. Hence an upper bound is given by the slowest of these rates, which establishes the claim for the rate of consistent estimation of $\sigma_a^2$. The respective upper bounds on the rate for estimating $\sigma_g^2$ and $\sigma_w^2$ follow from analogous arguments\qed


\subsection*{Proof of Theorem \ref{bootstrap_cons_thm}}

For bootstrap consistency it suffices to verify whether the limiting distributions of $r_{NT}^{-1}(\bar{Y}_{NT}^*-\bar{Y}_{NT})$ and $r_{NT}^{-1}(\bar{Y}_{NT}-\mathbb{E}[Y_{it}])$ coincide. 

For part (a), note that by Lemma \ref{var_comp_rate_lem} (a), $\tilde{\lambda}$ is (pointwise) consistent for $\lambda$ and that furhtermore the local parameter with $q_a+q_g>0$ and $q_v>0$ can only be achieved at drifting sequences, so that this case is irrelevant for point-wise convergence. For the remaining cases, the limiting distribution simplifies to
\[r_{NT}^{-1}(\bar{Y}_{NT}^*-\bar{Y}_{NT})\rightsquigarrow\left\{\begin{array}{lcl} Z_0^*&\hspace{0.5cm}&\textnormal{if }q_a+q_g>0\\ \sqrt{q_e}Z_0^* + \sqrt{q_v}V^*&&\textnormal{if }q_a+q_g=0
\end{array}\right.\]
where $Z_0^*\sim N(0,1)$, independent of $V:=\sum_{k=1}^{\infty}\frac{c_k}{\sigma_v}Z_{1k}Z_{2k}$ for independent standard normal random variables $Z_{11},Z_{21},\dots$. In particular, that distribution is equal to the asymptotic distribution of $r_{NT}^{-1}(\bar{Y}_{NT}^*-\mathbb{E}[Y_{it}])$. Hence claim (a) follows from Theorems \ref{sample_clt_thm} and \ref{bootstrap_clt_thm} and the triangle inequality.

For part(b), note that by Lemma \ref{var_comp_rate_lem} (a), $\hat{\lambda}$ is uniformly consistent if $q_v=0$, and that for every $K<\infty$, the random vector $\hat{Z}_{NT,K}$ is finite-dimensional. We can therefore adapt an argument by \cite{Agu10},\footnote{See the proof of their Theorem 1 for details.} to verify that it is sufficient to consider convergent subsequences for which the appropriately normalized parameters converge to proper limits.

Under such sequences, we can conclude from Theorems \ref{sample_clt_thm} and \ref{bootstrap_clt_thm} that the limits of the sampling and the bootstrap distribution coincide. Part (b) then follows from the triangle inequality \qed

\subsection*{Proof of Theorem \ref{refinement_thm}.}

We can establish the refinements of this bootstrap procedure by verifying the conditions for part (ii) of the main theorem in chapter 5 of \cite{Mam92}.


First note that the third moment of $\hat{a}_i$ under the sampling distribution is
\[\mathbb{E}[\hat{a}_i^3] = \left(\mathbb{E}[a_i^3] + \frac2T\mathbb{E}[a_iw_{it}^2] + \frac1{T^2}\mathbb{E}[w_{it}^3]\right)(1+O(1/N))\]
where we used the fact that $w_{it}$ is mean-independent of $a_i$. By the assumptions of the theorem and a central limit theorem, we then have $\frac1N\sum_{i=1}^N(\hat{a}_i^3-\mathbb{E}[a_i^3])=O_P(n^{-1/2})$. Hence, by standard calculations,
\begin{eqnarray}\nonumber \mathbb{E}_{NT}^*\left[\left(\hat{Z}_N^{a,*}\right)^3\right]
-\mathbb{E}\left[\left(\hat{Z}_N^{a}\right)^3\right]
&=&N^{-1/2}\left(\mathbb{E}_{NT}^*[(a_i^*)^3]-\mathbb{E}[a_i^3]\right)\\
\nonumber&=&O_P(N^{-1})
\end{eqnarray}

This amounts to establishing condition $DIFF_T(3,C)$ in \cite{Mam92} for the process $\hat{Z}_N^{a,*}$. Verifying the conditions $DIFF_S(2)$ and $VAR(2)$ follows similar steps and is more standard. Note that by inspection of the expression for $\mathbb{E}[\hat{a}_i^3]$, the conclusion does not hold in general under arbitrary drifting sequences for the second and third moments of $a_i,w_{it}$. Using the same arguments, we can establish conditions $DIFF_T(3,C)$, $DIFF_S(2)$, and $VAR(2)$ for $\hat{Z}_T^{g,*}$ at the respective rates in $T$.

For the analogous results for the components $\hat{Z}_{NT}^{e,*},\hat{Z}_{Nk}^{\phi,*},\hat{Z}_{Tk}^{\psi,*}$, note that by assumption $\mathbb{E}[\omega_i^3]=\mathbb{E}[\omega_t^3]=1$ and the draws are independent, so that that $\mathbb{E}[(\omega_i\omega_t)^3]=1$. Hence, the third moments of $e_{it}^*,(\phi^k(\alpha_i))^*,(\psi^k(\gamma_t))^*$ under the bootstrap distribution also converge in probability to the third moments of $e_{it},\phi^k(\alpha_i),\psi^k(\gamma_t)$ under the sampling distribution, using standard arguments analogous to the previous case. In particular, conditions $DIFF_T(3,C)$, $DIFF_S(2,C)$ and $VAR(2)$ in \cite{Mam92} hold for $\hat{Z}_{NT}^{e,*},\hat{Z}_{Nk}^{\phi,*},\hat{Z}_{Tk}^{\psi,*}$ and all $k=1,\dots,K$ at the respective rates in $NT$, $N$, and $T$. Furthermore, convergence in each of finitely many components implies joint convergence of cumulants for all components of $\hat{Z}_{NT,K}^*$.


Since under $q_v>0$ the statistic is not asymptotically pivotal, in the following we only consider the case in which the contribution of $\hat{Z}_{Nk}^{\phi},\hat{Z}_{Tk}^{\psi}$ through the Wiener chaos component is asymptotically negligible. By construction, $\hat{Z}_N^a$ and $\hat{Z}_T^g$ and their bootstrap versions $\hat{Z}_N^{a,*}$ and $\hat{Z}_T^{g,*}$ are independent. Also, the components of $\hat{Z}_N^a,\hat{Z}_T^g,\hat{Z}_{NT}^{e}$ as well as their bootstrap analogs are asymptotically uncorrelated. For third cumulants of weighted sums of $\hat{Z}_N^a$ and $\hat{Z}_{NT}^e$ we also need to consider the moments
\[\mathbb{E}[\hat{a}_i\hat{w}_{it}^2] = \mathbb{E}[a_iw_{it}^2](1+O(1/N)) \]
where $\mathbb{E}_{NT}^*[a_i^*(w_{it}^*)^2]-\mathbb{E}[\hat{a}_i\hat{w}_{it}^2]=O_P(N^{-1/2})$ by standard arguments. By similar arguments as for the third moments of $a_i$ and $g_t$, for any weights $s_1,s_2\geq0$, we then have
\begin{eqnarray}\nonumber \mathbb{E}_{NT}^*\left[\left(s_1\hat{Z}_N^{a,*}+s_2\hat{Z}_{NT}^{e,*}\right)^3\right]
-\mathbb{E}\left[\left(s_1\hat{Z}_N^{a}+s_2\hat{Z}_{NT}^{e,*}\right)^3\right]
&=&O_P(N^{-1})
\end{eqnarray}
with the analogous conclusion for weighted sums of $\hat{Z}_T^g$ and $\hat{Z}_{NT}^e$ and their bootstrap analogs.

Since $\hat{\lambda}\stackrel{p}{\rightarrow}\lambda$, we can combine convergence of the cumulants of the joint distribution of the individual components to verify that the conditions $DIFF_T(3,C)$, $DIFF_S(2)$, and $VAR(2)$ also hold for the weighted sums with rates in $N$ if $\sigma_a>0$ ($T$, respectively, if $\sigma_g>0$), or $NT$ if $\sigma_a=\sigma_g=0$ and $\sigma_e>0$, so that the conclusion follows from the main theorem in chapter 5 of \cite{Mam92}\qed

\subsection*{Proof of Theorem \ref{nonexhaust_cons_thm}:} The main arguments from the Proof of Theorem \ref{bootstrap_cons_thm} hold after a few minor modifications of the arguments for the case $q_v=0$. The only major complication arises if the second-order projection term $\frac1{NT\bar{p}^2}\sum_{i=1}^N\sum_{t=1}^TW_{it}[h(\alpha_i,\gamma_t)-a_i-g_t + \mathbb{E}[Y_{it}])$ remains relevant in the limit. In that case, the terms $\frac1{NT\bar{p}}\sum_{i=1}^N\sum_{t=1}^TW_{it}\phi_k(\alpha_i)\psi_k(\gamma_t)$ of the sparse representation can in general no longer be represented in terms of separate sample averages of $\phi_k(\alpha_i)$ and $\psi_k(\gamma_t)$, respectively.

We first consider the case of dyadic data, where the components of the second-order projection term takes the form
\[Q_k:=\frac1{N^2\bar{p}}\sum_{i=1}^N\sum_{j=1}^NW_{it}\phi_k(\alpha_i)\phi_k(\alpha_j)
\equiv\frac1{N^2\bar{p}}\phi_k'W\phi_k=\frac1{2N^2\bar{p}}\phi_k'(W+W')\phi_k\]
 for the vector $\phi_k:=(\phi_k(\alpha_1),\dots,\phi_k(\alpha_N))'$. To characterize the limit distribution for $N\sqrt{p}Q_k$, let $Z_k\sim N(0,I_N)$ and $\tilde{Q}_k:=\frac1{2N^2\bar{p}}Z_k'(W+W')Z_k$. Conditions for convergence of $N\sqrt{p}Q_k$ to $N\sqrt{p}\tilde{Q}_k$ were given by \cite{GTi99}. 

Now, by Assumption \ref{nonexhaustive_ass} (a), we either have that $\sup_{i=1,\dots,N}p_i\rightarrow0$, or that $\lim_N\bar{p}>0$. Hence we only need to distinguish two cases regarding the asymptotic behavior of $p_i$. For the first case with $\sup_{i=1,\dots,N}p_i\rightarrow0$, Corollary 2 in \cite{GTi99} implies that
\[\varrho(N\sqrt{p}Q_k,N\sqrt{p}\tilde{Q}_k)\leq(\mathbb{E}|\phi_k(\alpha_i)|^3)^2\sup_{i=1,\dots,N}\sqrt{p_i}\]
where $\varrho(X,Y):=\sup_x|F_X(x)-F_Y(x)|$ for any two random variables $X,Y$ with respective c.d.f.s $F_X$ and $F_Y$. Furthermore, in this case the asymptotic distribution of $N\sqrt{p}Q_k$ is Gaussian. By an analogous argument, we also find that the distribution of the bootstrap analog $N\sqrt{p}Q_k^*$ converges to $N\sqrt{p}\tilde{Q}_k$, so that bootstrap consistency follows from the triangle inequality. For the second case with $\bar{p}$ bounded away from zero, $p_i$ is bounded away from zero by a constant for at least two distinct units in $\{1,\dots,N\}$. In that case, consistency follows instead from Theorem 3 in \cite{GTi99}.

An extension to multilinear forms for the case in which each dimension of the random array corresponds to a different type of sampling unit can be obtained in a straightforward manner after stacking the random variates $\phi_k(\alpha_1),\dots,\phi_k(\alpha_N),\psi_k(\gamma_1),\dots,\psi_k(\gamma_T)$ and considering the symmetric quadratic form corresponding to the $(N+T)\times(N+T)$ matrix $A=\frac12[0,W;W'0]$ \qed

\normalsize
\clearpage

\bibliographystyle{econometrica}
\bibliography{mybibnew}

\end{document}